\documentclass[11pt]{article}

\usepackage{jheppub}
\usepackage{appendix}
\usepackage{graphicx}
\usepackage[export]{adjustbox}
\usepackage{subcaption}
\usepackage{color}
\usepackage{tikz}
\usepackage{tikz-cd}
\usepackage{slashed}
\usepackage{braket}
\usepackage{comment}
\newcommand{\ol}{\overline}
\newcommand{\wt}{\widetilde}
\newcommand{\Tr}{\mathrm{Tr}}

\newcommand{\pd}{{\partial}}

\newcommand{\C}{\mathbb{C}}
\newcommand{\R}{\mathbb{R}}
\newcommand{\Z}{\mathbb{Z}}

\newcommand{\PP}{\mathbb{P}}
\newcommand{\PT}{\mathbb{PT}}

\newcommand{\fsl}{\mathfrak{sl}}

\newcommand{\fg}{\mathfrak{g}}
\newcommand{\fh}{\mathfrak{h}}

\newcommand{\CA}{{\mathcal A}}
\newcommand{\CB}{{\mathcal B}}
\newcommand{\CC}{{\mathcal C}}

\newcommand{\CE}{{\mathcal E}}
\newcommand{\CF}{{\mathcal F}}
\newcommand{\CG}{{\mathcal G}}

\newcommand{\CI}{{\mathcal I}}

\newcommand{\CK}{{\mathcal K}}

\newcommand{\CO}{{\mathcal O}}

\newcommand{\CQ}{{\mathcal Q}}

\newcommand{\CX}{{\mathcal X}}
\newcommand{\CY}{{\mathcal Y}}
\newcommand{\CZ}{{\mathcal Z}}

\newcommand{\be}{\begin{equation}}
	\newcommand{\ee}{\end{equation}}

\title{Scattering off of Twistorial Line Defects}
\author[1]{Niklas Garner}
\emailAdd{nkgarner@uw.edu}
\author[1]{Natalie M. Paquette}
\emailAdd{npaquett@uw.edu}
\affiliation[1]{Department of Physics, University of Washington, Seattle, USA}

\abstract{The recently devised chiral algebra bootstrap computes the form factors of a special class of ``twistorial'' 4d QFTs as correlation functions of the theory's 2d celestial chiral algebra.
Examples of twistorial theories include self-dual Yang-Mills theory coupled to special massless matter content, and certain form factors in these theories are equivalent to a subset of MHV amplitudes in massless QCD,  coupled to the same matter.
In this paper, we extend the chiral algebra bootstrap to include scattering in the presence of charged sources, using a self-dual dyon in a twistorial theory as our main example. Self-dual theories in the presence of such sources lift to holomorphic gauge theories on non-Hausdorff twistor space, and we generalize the Koszul duality construction of Costello and Paquette to this setting.
With this approach, we easily reproduce a recent formula of Adamo, Bogna, Mason, and Sharma for $n$-point MHV scattering of gluons off the self-dual dyon.}

\begin{document}
\maketitle

\section{Introduction \& Summary}

Scattering massless particles off of magnetic monopoles is an old problem with famous applications to baryon decay \cite{rubakov1988monopole, callan1983monopole}, yet leads to puzzling features, such as the appearance of fractional quantum numbers, unfamiliar from standard S-matrix lore \cite{Rubakov:1982fp}.
This is a challenging problem to study directly, in part because there is no mutually local Lorentz invariant Lagrangian description of both electrically and magnetically charged fields to which we can apply our usual off-shell perturbative techniques.
Recently there have been a variety of novel approaches to and perspectives on this problem (e.g. \cite{Csaki:2021ozp, Csaki:2022tvb, Hamada:2022eiv, Brennan:2023tae, brennan2023callan, vanBeest:2023dbu, vanBeest:2023mbs, Mouland:2024zgk}), including a revisiting of on-shell methods and little group symmetries \cite{Csaki:2021ozp} and a proposal that the outgoing states are in a different Hilbert space containing states dressed by topological defects for generalized symmetries \cite{vanBeest:2023dbu}.%
\footnote{Our analysis is consistent with the appearance of pairwise helicity described in \cite{Csaki:2021ozp, Csaki:2022tvb}, as is the work of Brennan \cite{brennan2023callan}.
We believe that it would be useful and interesting to understand how the Callan-Rubakov effect is manifested in self-dual/twistorial gauge theories, but we defer this to future work.} %

Although non-abelian gauge theories admit smooth field configurations with magnetic charge, which realize (potentially massless) genuine states of the theory, we will focus in this paper on the scenario in which the magnetically charged particles are infinitely heavy, \emph{i.e.} we study a magnetically charged source in the probe limit, which can therefore be modeled as an 't Hooft-like line defect extended in (Euclidean) time.
More generally, it can be useful to study scattering in the presence of other heavy probe particles or, equivalently, line defects.
Dually, one can consider scattering in the presence of the non-trivial backgrounds sourced by these defects.
Famously, Wilson lines in non-abelian gauge theory can be modeled by the insertion of heavy quarks, and this description features in heavy quark effective theory; in condensed matter physics, the Kondo problem studies electron scattering in the presence of a magnetic impurity; and understanding scattering in the presence of general QCD backgrounds plays an important role in describing the processes observed in heavy ion collisions.
In each of these problems, it is incredibly difficult to describe low-multiplicity amplitudes and all-multiplicity amplitudes are unheard of.

A simplified setting for studying these scattering problems is that of 4d twistorial theories.
Twistorial theories admit local holomorphic lifts to twistor space \cite{Costello:2021bah, CP22} and enjoy many properties such as strong analytic constraints on correlation functions, are scale invariant, hence UV complete, and possess quantum integrable features as a consequence of this lift.
The main aim of the present paper is to develop tools for studying these scattering problems in the context of the celestial chiral algebra bootstrap program initiated by the second author with K. Costello.
We view this as a necessary first step towards understanding more refined observables in involving defects.
Examples of such observables are the cusp anomalous dimension which involves studying insertions of multiple line defects meeting at a point; we intend to describe these observables in follow-up work.
Of course, a general line defect or background will spoil these remarkable features of twistorial field theories.
Instead, we focus on a special class of line defects that admit a compatible lift to twistor space -- we call such a defect \emph{twistorial}.
It is natural to expect that twistorial line defects preserve many of the desirable features of twistorial field theories; we will see this is the case in an explicit example.
As we will see, focusing on this class of theories and defects/backgrounds will allow us to access all-multiplicity scattering amplitudes in a relatively simple manner by way of algebra.

The starting point for our analysis is the twistorial uplift of self-dual field configurations in the presence of a line-like source.
Massless fields in spacetime with a source along some analytic worldline can be lifted to twistor space \cite{bailey1985twistors, Bailey:1989if}, see also Chapter 6 of \cite{MH1990}, and so it is natural to study them with the tools available to twistorial theories.
The prototypical example of a twistorial line defect is the insertion of a probe dyon with self-dual electric and magnetic charges into a self-dual twistorial gauge theory.%
\footnote{Twistorial theories require that local gauge anomalies of their holomorphic uplift to twistor space vanish \cite{Costello:2021bah}, which requires the introduction of special massless matter content; by cohomological arguments \cite{Costello:2019jsy}, it is believed that there are no other possible perturbative inconsistencies.} %
This self-dual dyonic probe can be viewed as a self-dual combination of Wilson and 't Hooft lines and the fields it sources admit several descriptions on twistor space, see \emph{e.g.} \cite{PS79, PS90}.
For twistorial self-dual gauge theories, these descriptions are expected to be consistent quantum mechanically, at least in perturbation theory.
These twistorial methods have recently been used in \cite{ABMS1} to obtain simple, closed-form expressions for tree-level 2-point MHV scattering amplitudes in interesting self-dual backgrounds, such as 2-point scattering off of self-dual dyons and Taub-NUT geometries in self-dual gauge theories and gravity, and equally simple all-multiplicity expressions for tree-level MHV scattering amplitude of gluons in the presence of a self-dual dyon in \cite{ABMS2}.
(See also \cite{Adamo:2020yzi, Adamo:2022mev, ABZ23} for a study of scattering in the presence of self-dual \textit{radiative} backgrounds.)

Following the work of \cite{CP22}, it is also known that form factors in 4d twistorial theories on flat spacetime can be reformulated as correlation functions in a 2d chiral algebra supported on the twistor sphere above the origin (equivalently, the celestial sphere); this chiral algebra may be thought of as the celestial chiral algebra of the 4d theory \cite{Guevara:2021abz, Strominger:2021mtt}.
When the twistorial uplift is gauge-anomaly-free, there is even a(n ordinary, local) chiral algebra at loop-level.
We refer to \cite{CP22} for a thorough discussion and review of this program and \cite{Costello:2022upu, Costello:2023vyy, Dixon:2024mzh}, which use the chiral algebra bootstrap to compute $n$-point one- and two-loop amplitudes, with special external helicity configurations, in QCD coupled to a ``twistorial amount'' of fundamental matter.%
\footnote{The chiral algebra bootstrap is rational, hence UV and IR finite, and quantum-mechanically consistent for twistorial theories.
The regulator implicit in this program is still not well understood, but work of Dixon-Morales \cite{Dixon:2024mzh} can reproduce these answers using a simple mass regulator.
This regulator produces Feynman integrals that are notably simpler than, say, dimensional regularization and understanding its properties is itself a subject of ongoing work.}%

The proofs of the chiral algebra bootstrap in \cite{CP22}, which include an application of Koszul duality \cite{CP20, PW21}, are formulated on the twistor space of flat spacetime.
The latter can be viewed as the total space of a fibration over the Riemann sphere, $\PT = \text{Tot}(\CO(1)^{\oplus 2} \to \PP^1)$.
In this note, we show how to generalize the chiral algebra bootstrap to the twistor spaces that arise from the presence of line-like sources.
To account for the multi-valuedness of spacetime fields upon complexifying the complement of the source, the corresponding twistor spaces are non-Hausdorff \cite{bailey1985twistors} and the Penrose transform, which identifies Dolbeault cohomology classes on twistor space with zero-rest-mass states in spacetime, must incorporate a form of relative cohomology \cite{Bailey90}.
We explain how to obtain the celestial chiral algebra OPEs from Koszul duality on such twistor spaces, and discuss how the conformal blocks of the chiral algebra on the celestial sphere%
\footnote{Note that our celestial chiral algebras are nonunitary, and admit an infinite number of conformal blocks even on the sphere.} %
may be obtained from holomorphic Wilson lines.

One surprising aspect of our analysis is that the celestial chiral algebra in the presence of a (twistorial) line defect is generally equipped with a differential: it is a differential graded Lie algebra (DGLA).
From a twistorial perspective, the presence of this differential is not at all surprising and stems from the non-Hausdorff nature of the underlying twistor space.
The DGLA of celestial symmetries is a concise way of encoding all of the symmetries as well as the constraints they must satisfy.
We find that the cohomology in degree 0 can be identified with a subalgebra of the ordinary celestial symmetry algebra; it is natural to view these symmetries as the ordinary celestial symmetries that preserve the dyonic background.
Nonetheless, there is higher cohomology extending these ordinary celestial symmetries.

We use the self-dual dyon as a running example to illustrate how to use the tools we develop in this paper.
In the last section we reproduce the $n$-point tree-level MHV amplitudes of \cite{ABMS2} using the chiral algebra bootstrap.
As always, one virtue of the chiral algebra bootstrap is that, given the two-point function (such as those of \emph{e.g.} \cite{ABMS1}, which we derive from several perspectives in the text), the generalization to the $n$-point quantity is an exceedingly simple and brief induction argument.

There are many natural questions and generalizations we leave to future work.
We conclude with a brief sample.
\begin{enumerate}
\item It is a straightforward exercise for the reader to study tree-level scattering off of a self-dual dyon in other gauge theories by essentially repeating our Yang-Mills computation.%
\footnote{We note, however, that these gauge theories are generally not twistorial and so this computation cannot be performed beyond tree level.
For a genuine twistorial gauge theories, e.g. $SU(3)$ QCD with $N_f = 9$, our method is quantum-mechanically consistent.} %
Similar amplitudes for self-dual theories in the presence of other charged sources can also be readily obtained.\\
\item It would be natural to use the chiral algebra bootstrap to study loop-level scattering off of the self-dual dyon, and more general twistorial line defects, as well.
To do this, one must consider self-dual gauge theory coupled to additional anomaly-cancelling massless matter (in addition to the inclusion of the dyon).\\
\item  One would also like to compute scattering amplitudes for non-trivial gravitational backgrounds, such as the Taub-NUT scattering formulas of \cite{ABMS1}, using the chiral algebra bootstrap.
There, as in the case of flat spacetime, we need a better understanding of what the 4d theory is computing, since there are no local gauge-invariant operators in gravity.
It is expected, and also suggested in \cite{Bittleston:2022jeq}, that one should consider (holomorphic descent on) operators with nonzero ghost number, or (relatedly) study an observable encoding deformations of the self-dual gravity theory.\\
\item One way to formulate our celestial chiral algebras is as the algebra of boundary local operators in a 3d holomorphic topological theory; this arises from a certain three-sphere reduction in twistor space \cite{CP22}.
This point of view is convenient for understanding various aspects of the chiral algebra bootstrap correspondence, such as the space of conformal blocks.
In \cite{Garner:2023izn}, for example, the present authors showed that boundary monopoles of the 3d theory%
\footnote{These are not to be confused with the 4d dyons that are the focus of the present work.} %
naturally extend the celestial chiral algebra by certain spectral flow modules; these 3d monopoles have a straightforward interpretation in 6d as perturbing around a non-trivial background field configuration, but their 4d spacetime interpretation remains puzzling.
One should be able to formulate a similar 3d picture from dimensional reduction of the non-Hausdorff twistor spaces of this note. It would be interesting if the 3d perspective sheds more light on this system.\\
\item It would be very interesting if our chiral algebra could be connected to (a self-dual version of) the arguments in \cite{vanBeest:2023dbu}, which studies an effective two-dimensional CFT formulation of the scattering problem.
There, the outgoing states are argued to live in a twisted Hilbert space, in the 2d CFT sense, and it would be interesting to find signatures of that in this setting.  \\
\item Finally, it could be fruitful to generalize this formalism to scattering around other self-dual backgrounds.
This line of inquiry is related to the top-down holographic models of \cite{CPS1, CPS2} (see also \cite{Bittleston:2023bzp}).
In these examples, the curved (but still asymptotically flat) background geometry sets a scale for the problem which manifests in the chiral algebras deforming and, in particular, acquiring a nonzero central charge.
Conjectures regarding which self-dual backgrounds deform the celestial chiral algebra have been made in \cite{ABZ23}.
Our analysis in Section \ref{sec:selfdualdyon} of non-abelian gauge theory on the non-Hausdorff twistor space mentioned above suggests an avenue for describing and studying non-abelian analogues of the self-dual dyon in terms of (equivariant) coherent sheaves on the affine Grassmanian as in the geometric Langlands program \cite{Kapustin:2006pk}.
\end{enumerate}

\section{Line Defects and Twistor Space}
We are interested in understanding scattering in 4d twistorial field theories with line defects, whose support we denote by $\ell$.
We can dually describe a line defect by the field configuration on $\R^4 \backslash \ell$ it sources.
There is a natural twistor correspondence for self-dual fields on $\R^4 \backslash \ell$ that we describe below; see Chapter 6 of \cite{MH1990} for more details.
Our main interest will be those special line defects that arise via this twistor correspondence; we call them \emph{twistorial} line defects.

We start by reviewing some notation.
We denote homogeneous coordinates on twistor space $\PT = \text{Tot}(\CO(1)^{\oplus 2} \to \PP^1)$ by $[\lambda_\alpha, \mu^{\dot{\alpha}}]$, where the $\lambda_\alpha$ do not vanish simultaneously and realize coordinates on the base $\PP^1$.
We cover $\PT$ with two affine open sets as follows.
When $\lambda_1 \neq 0$ we set
\be
\label{eq:patch1}
	z = \frac{\lambda_2}{\lambda_1} \qquad v^{\dot{\alpha}} = \frac{\mu^{\dot{\alpha}}}{\lambda_1}
\ee
and, similarly, when $\lambda_2 \neq 0$ we set
\be
\label{eq:patch2}
	w = \frac{\lambda_1}{\lambda_2} \qquad u^{\dot{\alpha}} = \frac{\mu^{\dot{\alpha}}}{\lambda_2}\,.
\ee
On the overlap where both $\lambda_\alpha$ are nonzero or, equivalently, $z, w \neq 0$, these coordinates are related as
\be
	w = z^{-1} \qquad u^{\dot{\alpha}} = z^{-1} v^{\dot{\alpha}}
\ee
and
\be 
	z = w^{-1} \qquad v^{\dot{\alpha}} = w^{-1} u^{\dot{\alpha}}
\ee

Every point $x^a \in \C^4$ in complexified spacetime has associated to it a (linearly) embedded $\PP^1$ (denoted $\PP^1_x$) given by the following incidence relation:
\be
	\mu^{\dot{\alpha}} = x^{\alpha \dot{\alpha}} \lambda_\alpha \qquad x^{\alpha \dot{\alpha}} := \begin{pmatrix}
		x^1 + i x^2 & x^3 + i x^4\\
		x^3 - i x^4 & -x^1 + i x^2\\
	\end{pmatrix}
\ee
Explicitly, in the affine patch with coordinates $z, v^{\dot{\alpha}}$ these relations read
\be
	v^{\dot{1}} = x^1 + i x^2 + z(x^3-i x^4) \qquad \textnormal{and} \qquad v^{\dot{2}} = x^3 + i x^4 - z(x^1 - i x^2)\,.
\ee

\subsection{Sources from non-Hausdorff twistor spaces and relative cohomology}
\label{sec:sources}
A particularly nice class of line defects in a 4d twistorial QFT will be those that descend from their parent 6d holomorphic theory via a suitable analogue of the twistor correspondence for sourced fields.
For concreteness, we consider defects in Euclidean $\R^4$ along the line $\ell = \{x^1 = x^2 = x^3 = 0\}$.
Fields defined on the complement of $\ell$ need not be single-valued upon complexifying spacetime, essentially due to the fact that $r = \sqrt{(x^1)^2 + (x^2)^2 + (x^3)^2}$ is double-valued.
In fact, due to Hartog's lemma, any field that is single-valued on the complexification of $\R^4 \backslash \ell$ necessarily extends over $\ell$; a general field becomes double-valued upon complexification.
The twistor correspondence thus relates zero-rest-mass fields on $\R^4 \backslash \ell$ to holomorphic fields on the twistor space of the \emph{double cover} (of the complexification) thereof.

The twistor space of this double cover and the resulting twistor correspondence are due to Bailey \cite{bailey1985twistors}.
First, note that the above incidence relation implies that the points on any twistor sphere above (the complexification of) $\ell$ satisfy $v^{\dot{1}} = -z v^{\dot{2}}$.
In particular, the union of these twistor spheres is identified with the quadric defined by the vanishing of $Q = v^{\dot{1}} + z v^{\dot{2}}$.
Together with $\wt{Q} = w u^{\dot{1}} + u^{\dot{2}}$, $Q$ can be viewed as a distinguished section of $\CO(2) \to \PT$.
We then consider the non-Hausdorff space $\PT_Q$ obtained by gluing two copies of $\PT$ over $Q = 0$:
\be
	\PT_Q := \PT \cup_{Q \neq 0} \PT
\ee
This non-Hausdorff space has many similarities to the plane with a doubled origin or its algebraic analog the formal bubble/raviolo; this space is realized by gluing two copies of the complex plane (or formal disk) away from $z = 0$ and is commonplace in various aspects of twisted 3- and 4-dimensional quantum field theories including the geometric Langlands correspondence \cite{Kapustin:2006pk}.
We will distinguish these two copies of $\PT$ with $+$ and $-$.
This space can be covered by four open sets, essentially two copies of the above open sets with coordinates in Eqs. \eqref{eq:patch1} and \eqref{eq:patch2}, where we identify these open sets away from $Q = 0$.
Focusing on the patches over the northern hemispheres $\lambda_{1, \pm} \neq 0$, we glue two copies of $\C^3$ over $\C^3 \backslash \{Q = 0\}$ with coordinates identified as
\be
	z_+ = z_- \qquad v^{\dot{\alpha}}_+ = v^{\dot{\alpha}}_-\,.
\ee  
The southern hemispheres are treated identically.
The twistor correspondence relevant for fields sourced along $\ell$ describes these configurations as cohomology classes on $\PT_Q$ (suppressing the underlying sheaf):
\be
	\{\text{self-dual fields sourced along } \ell\} \longleftrightarrow H^{(0,1)}(\PT_Q)\,.
\ee

It is easy to describe these cohomology groups in terms of objects defined on ordinary twistor space $\PT$ \cite{Bailey90}.
We start from the Mayer-Vietoris sequence based on the cover of $\PT_Q$ by the two copies of $\PT$.
At the level of forms (suppressing the underlying sheaves), we get the following short exact sequence:
\be
	0 \to \Omega^{(0,n)}(\PT_Q) \overset{\alpha}{\to} \Omega^{(0,n)}(\PT) \oplus \Omega^{(0,n)}(\PT) \overset{\beta}{\to} \Omega^{(0,n)}(\PT \backslash \{Q = 0\}) \to 0
\ee
where $\alpha(f) = (f|_\PT, f|_\PT)$ and $\beta(f,g) = f|_{Q \neq 0} - g|_{Q \neq 0}$.
That this sequence is exact says that $\Omega^{(0,n)}(\PT_Q)$ can be identified with the kernel of $\beta$; in other words, an $n$-form on $\PT_Q$ is a pair of $n$-forms on $\PT$ $(f,g)$ that agree upon restriction to $Q \neq 0$.
To have a complete description of fields sourced along $\ell$, it is too restrictive to simply ask for $1$-forms $Z_\pm$ on the two copies of $\PT$ that agree away from $Q = 0$ and satisfy the equation of motion
\be
\label{eq:PTQcoho1}
	\ol{\pd} Z_\pm = 0\,,
\ee
modulo shifts by $\ol{\pd}$-exact 1-forms.
We should only ask that the difference between them is exact away from $Q = 0$.
In other words, there is a 0-form $Z_0$ defined away from $Q = 0$ that satisfies the equation of motion
\be
\label{eq:PTQcoho2}
	\ol{\pd} Z_0 + Z_+| - Z_-| = 0\,,
\ee
where $Z_\pm|$ denotes the restriction of $Z_\pm$ to $Q \neq 0$.
Under the above redundancy $Z_\pm \to Z_\pm + \ol{\pd} Y_\pm$ for 0-forms $Y_\pm$, $Z_0$ must transform as $Z_0 \to Z_0 + Y_-| - Y_+|$ to ensure its equations of motion are invariant.

These fields, their equations of motion, and the redundancies thereof can be neatly described cohomologically in terms of full forms as in the BV/BRST formalism.
Namely, we consider fields $\CZ_\pm$ valued in $\Omega^{(0,\bullet)}(\PT)[1]$ and a field $\CZ_0$ valued in $\Omega^{(0,\bullet)}(\PT\backslash\{Q=0\})$.
The $[1]$ appearing in $\Omega^{(0,\bullet)}(\PT)[1]$ indicates a shift in cohomological degree/ghost number so that the field $\CZ_\pm$ is homogeneous with ghost number 1, whereas $\CZ_0$ is homogeneous with ghost number 0; it is natural to give the coordinate $1$-forms $\text{d} \ol{z}$, $\text{d} \ol{v}^{\dot \alpha}$ ghost number 1.
The equations of motion for these fields and the redundancies thereof are encoded in the ghost number $1$ BV/BRST differential $\CQ$ given by
\be
	\CQ \CZ_\pm = \ol{\pd} \CZ_\pm\,, \qquad \CQ \CZ_0 = \ol{\pd} \CZ_0 + \CZ_+| - \CZ_-|\,.
\ee
We will often denote this collection of fields by $\widehat{\CZ}$.
The above discussion reappears upon decomposing these fields into their components of a given form degree as
\be
	\CZ_\pm = \sum_{i = 0}^3 \CZ^{(i)}_\pm \qquad \CZ_0 = \sum_{i = 0}^3 \CZ^{(i)}_0
\ee
so that the above differential acts on these components as
\be
	\CQ \CZ^{(i)}_\pm = \ol{\pd} \CZ^{(i-1)}_\pm \qquad \CQ \CZ_0^{(i)} = \ol{\pd} \CZ^{(i-1)}_0 + \CZ_+^{(i)}| - \CZ_-^{(i)}|
\ee
where we set $\CZ^{(-1)}_\pm = 0$ and $\CZ^{(-1)}_{0} = 0$.
The physical fields $Z_\pm$, $Z_0$ are identified with the component fields of ghost number $0$, \emph{i.e.} the 1-forms $\CZ^{(1)}_\pm$ and 0-form $\CZ^{(0)}$; their equations of motion are implemented cohomologically as the $\CQ$-variation of the anti-fields $\CZ^{(2)}_\pm$ and $\CZ^{(1)}_0$, whose components have ghost number $-1$.
Similarly, the 0-forms $\CZ^{(0)}_\pm$ have ghost number 1 and are interpreted as ghosts for the above shift redundancy.

It is useful to unpack this complex a bit so that we may connect to the traditional twistorial description of self-dual fields sourced on a worldline due to Bailey \cite{bailey1985twistors, Bailey:1989if}.
If we replace the fields $\CZ_\pm$ by their average $\CZ = \tfrac{1}{2}(\CZ_+ + \CZ_-)$ and their difference $\CX = \CZ_+ - \CZ_-$, the field $\CZ$ decouples from $\CX$ and $\CY = \CZ_0$:
\be
	\CQ \CZ = \ol{\pd} \CZ \qquad \CQ \CX = \ol{\pd} \CX \qquad \CQ \CY = \ol{\pd} \CY + \CX|
\ee
The cohomology of $\CQ$ acting on $\CZ$ clearly computes $H^{(0,\bullet)}(\PT)[1]$, whose ghost number 0 component is simply $H^{(0,\bullet)}(\PT)$ describing ordinary self-dual fields via the standard Penrose transform.
The cohomology of $\CQ$ acting on $(\CX, \CY)$, on the other hand, computes the (Dolbeault) cohomology (of $(0,\bullet)$-forms) with support in the closed subset $Q = 0$, denoted $H^{(0,\bullet)}_Q(\PT)[1]$.%
\footnote{This cohomology group is often denoted $H^{(0,\bullet)}_Z(\PT)$ for $Z = Z(Q)$ the vanishing locus of $Q$.
We use the present notation to align that of \cite{Bailey90}.} %
That this is true follows from the fact that the degree 0 cohomology is simply holomorphic functions on $\PT$ which vanish on $Q \neq 0$, \emph{i.e.} holomorphic functions supported on $Q = 0$:
\be
	\CQ \CX^{(1)} = \ol{\pd} \CX^{(0)} \sim 0 \qquad \CQ \CY^{(0)} = \CX^{(0)}| \sim 0\,.
\ee
In degree $n$, the equations of motion for $\CX^{(n)}$ say it is closed with exact restriction to $Q \neq 0$
\be
	\CQ \CX^{(n+1)} = \ol{\pd} \CX^{(n)} \sim 0 \qquad \CQ \CY^{(n)} = \ol{\pd}\CY^{(n-1)} + \CX^{(n)}| \sim 0
\ee
and gauge transformations shift it by exact forms (with the form realizing its exactness on $Q \neq 0$ transforming accordingly)
\be
	\CQ \CX^{(n)} = \ol{\pd} \CX^{(n-1)} \qquad \CQ \CY^{(n-1)} = \ol{\pd} \CY^{(n-2)} + \CX^{(n-1)}|.
\ee

Putting this together, we see that the cohomology groups $H^{(0,\bullet)}(\PT_Q)[1]$ take the following form \cite{Bailey90}:
\be
	H^{(0,\bullet)}(\PT_Q)[1] \simeq H^{(0,\bullet)}(\PT)[1] \oplus H^{(0,\bullet)}_Q(\PT)[1]
\ee
Identifying the physical fields with form-degree 1 cohomology classes $H^{(0,1)}(\PT_Q)$, we see that the first term gives field configurations that are source-free via the usual twistor correspondence.
The cohomology group $H^{(0,1)}_Q(\PT)$ describes the allowed structure of sourced fields.
We note that this is compatible with the classic results \cite{bailey1985twistors, Bailey:1989if} relating sourced fields to cohomology classes  \emph{relative} to $Q \neq 0$, as the two notions are equivalent: the above model is simply a Dolbeault analog of relative de Rham cohomology \cite{BottTu}.

Going forward, we will drop the vertical line $|$ indicating a restriction to $Q \neq 0$; we will generally use the notation ${}_\pm$ and ${}_0$ to denote fields coming from each copy of $\PT$ and from the overlapping region $Q \neq 0$, and expressions involving fields with ${}_0$ should be understood to take place in the region with $Q \neq 0$.

\subsection{From \texorpdfstring{$\PT_Q$}{PTQ} to spacetime}
\label{sec:reduction}
We now describe how to turn the above cohomology classes into fields on spacetime.
As we will see, for each point $x \in \R^4 \backslash \ell$ there is a distinguished $\PP^1$ in $\PT_Q$; we will obtain the value of a given spacetime field at $x$ by integration over this $\PP^1$.

To identify the desired $\PP^1$, we note that the twistor sphere $\PP^1_x \subset \PT$ for $x \in \R^4 \backslash \ell$ intersects $Q = 0$ at the antipodal points $z = \zeta$ and $z = -\ol{\zeta}^{-1}$:
\be
	Q|_{x} := Q|_{\PP^1_x} = \frac{-2r(z - \zeta)(z\ol{\zeta} + 1)}{1+|\zeta|^2}
\ee
where
\be
	r = \sqrt{(x^1)^2 + (x^2)^2 + (x^3)^2} \qquad	\zeta = \frac{x^1 + i x^2}{r - x^3} \qquad \ol{\zeta} = \frac{x^1 - i x^2}{r - x^3}
\ee
We can lift this $\PP^1$ to $\PT_Q$ so that each these intersections happen on different copies of $\PT$ inside $\PT_Q$: for $z \neq -\ol{\zeta}^{-1}$ (resp. $z \neq \zeta$) we lift the point $(x, z) \in \PT$ to the first (resp. second) copy of $\PT$.
The points $(x,z)$ with $z \neq \zeta$ and $z \neq - \overline{\zeta}^{-1}$, and hence with $Q \neq 0$, do not get lifted to multiple points as the two prescriptions send it to points that are identified.
We will denote this copy of $\PP^1$ inside $\PT_Q$ by $\PP^1_{Q,x}$.

In order to obtain a non-zero answer, we need to integrate over $\PP^1_{Q,x}$ a section of $\CO(-2)$.
For simplicity, assume $\CZ_\pm$ and $\CZ_0$ are all valued in $\CO(-2)$, corresponding to the case of a scalar field in 4d; other helicities can be treated as in the usual twistor correspondence by introducing factors of $\lambda_{\alpha, \pm}$ or taking derivatives with respect to $\mu^{\dot \alpha}$.
Given such a triple, we consider the following integral:
\be
\begin{aligned}
	Z_{\gamma}(x) & = \int_{\PP^1_{Q,x}, \gamma} \widehat{\CZ} := \int_{D_+} \CZ_+ + \int_{D_-} \CZ_- + \oint_\gamma \CZ_0\\
	& = \int_{D_+} \CZ_+^{(1)} + \int_{D_-} \CZ_-^{(1)} + \oint_\gamma \CZ_0^{(0)}
\end{aligned}
\ee
where $\gamma$ is a curve wrapping once around $\PP^1_{Q,x}$ and separating $\zeta$ and $-\ol{\zeta}^{-1}$, with $D_{\pm}$ the two resulting halves of $\PP^1_{Q,x}$ satisfying $\pd D_\pm = \pm\gamma$.

It is easy to see that this expression is gauge invariant due to Stokes' theorem:
\be
	\CQ Z_{\gamma}(x) = -\int_{D_+} \ol{\pd} \CZ_+ - \int_{D_-} \ol{\pd} \CZ_- + \oint_\gamma \big(\CZ_+ - \CZ_-\big) = 0
\ee
Moreover, homologous choices of $\gamma$ lead to cohomologous 4d field configurations:
\be
\begin{aligned}
	Z_\gamma(x) - Z_{\gamma'}(x) & = \bigg(\int_{D_+} \CZ_+ + \int_{D_-} \CZ_- + \oint_{\gamma} \CZ_0\bigg) - \bigg(\int_{D_+'} \CZ_+ + \int_{D_-'} \CZ_- + \oint_{\gamma'} \CZ_0\bigg)\\
	& = \int_{D_+ - D_+'}\CZ_+ -  \CZ_- + \ol{\pd} \CZ_0 = \CQ \bigg(-\int_{D_+ - D_+'} \CZ_0\bigg) \sim 0
\end{aligned}
\ee
Note that this integral reproduces the usual twistor correspondence when evaluated on field configurations with $\CZ_+ = \CZ_-$ and $\CZ_0 = 0$.

\subsubsection{Example: sourced scalar field}
\label{sec:scalar}

The simplest example of a sourced field is that of the scalar field profile $Z_{4d} = r^{-1}$, which solves the massless wave equation on the complement of $r = 0$.
The twistorial uplift of this configuration can be described using the above language: we consider the field configuration
\be
	\CZ_\pm = 0 \qquad \CZ_0 =\frac{\text{d}z}{2 \pi i} \big(-2 Q^{-1}\big)\,.
\ee
which clearly solves the above equations of motion.
The resulting 4d field configuration is thus obtained by taking the residue of $Q^{-1}$ at $z = \zeta$ (or at $z = - \ol{\zeta}$)
\be
	Z_{\gamma}(x) = \oint_\gamma\frac{\text{d}z}{2\pi i} \big(-2 Q^{-1}\big) = \frac{2}{(x^1 - i x^2)(\zeta+\ol{\zeta}^{-1})} = \frac{1}{r}
\ee

\subsubsection{Example: anti-self-dual dyon}

The next simplest example is that of an anti-self-dual dyonic field.
This should arise for a suitable field configuration where the fields $\CZ_\pm$ and $\CZ_0$ are sections of $\CO(-4)$.
To reproduce the anti-self-dual dyon, we consider the field configuration
\be
	\CZ_\pm = 0 \qquad \CZ_0 = \frac{\text{d}z}{2 \pi i} \big(-Q^{-2}\big)
\ee
We can compute the resulting 4d field configuration as above, but now including a factor of $\lambda_\alpha \lambda_\beta$ so that our integrand has the correct homogeneity:
\be
	Z_{\gamma}(x) = \oint_\gamma \frac{\text{d}z}{2\pi i} \begin{pmatrix} z^2 & z\\ z & 1\end{pmatrix} \big(-Q^{-2}\big) = \begin{pmatrix}
		\frac{x^1 + i x^2}{r^3} & \frac{-x^3}{r^3}\\
		\frac{-x^3}{r^3} & \frac{-x^1 + i x^2}{r^3}
	\end{pmatrix}
\ee

\section{The Self-dual Dyon from \texorpdfstring{$\PT_Q$}{PTQ}}
\label{sec:selfdualdyon}
We reviewed in the previous section how self-dual fields sourced on the worldline $\ell \subset \R^4$ can be described in terms of cohomology classes on the non-Hausdorff twistor space $\PT_Q$.
In this section we describe the prototypical example of a twistorial line defect in self-dual gauge theories: the self-dual dyon.
This defect is merely a self-dual combination of Wilson and 't Hooft lines.
Although this line operator sources a self-dual background, it has many interesting features due to the topologically non-trivial nature of the gauge fields coming from the dyon's magnetic charge.
We embed the self-dual dyon into non-abelian gauge theory in the usual manner.
We let $G$ be a reductive complex Lie group, the complexification of the 4d (compact) gauge group $G_c$, with (complex) Lie algebra $\fg$.
Fixing a maximal torus $T \subset G$ with Cartan subalgebra $\fh \subset \fg$, we choose a cocharacter $\nu: U(1) \to T$ of $T$, which we can identify with a semisimple element $T_0 \in \fh$.
We choose a basis $T_a$ of weight vectors in $\fg$ with weights $e_a$; the assumption that this is a cocharacter ensures the $e_a$ are integral.

To understand perturbation theory around this dyonic background, we realize self-dual gauge theory twistorially as holomorphic BF theory on twistor space:
\be
	S_{4d} = \int_{\R^4}\Tr(B F(A)_-) \rightsquigarrow S_{6d} = \int_{\PT} \Tr(\CB \CF^{(0,2)}(\CA))
\ee
where $\CF^{(0,2)}(\CA)$ is the curvature of the $(0,1)$ gauge field $\CA$ and $\CB$ is an adjoint-valued section of $\CO(-4)$.
We can then couple 4d self-dual gauge theory to the self-dual dyon by working around a certain abelian field configuration, called the Coulomb bundle $\CC$, embedded into $G$ along $T_0$.

\subsection{Holomorphic gauge theory on \texorpdfstring{$\PT_Q$}{PTQ}}

We start by formulating holomorphic gauge theory on $\PT_Q$, again in the BV/BRST formalism.
This theory reduces to holomorphic $BF$ theory on each copy of $\PT$, whose equations of motion are encoded in the differential $\CQ$ as
\be
	\CQ \CA_\pm = \ol{\pd} \CA_\pm + \tfrac{1}{2}[\CA_\pm, \CA_\pm]\,, \qquad \CQ \CB_\pm = \ol{\pd} \CB_\pm + [\CA_\pm, \CB_\pm]\,.
\ee
We would then like to identify these fields away from $Q = 0$, but it is too much to ask that they are equal -- we only ask that they differ by a gauge transformation.
We can impose this condition as the equations of motion for fields $\CG, \CB_0$:
\be
	\CQ \CG = \ol{\pd} \CG + \CA_+ \CG - \CG \CA_-\,, \qquad \CQ \CB_0 = \ol{\pd} \CB_0 + [\CA_+, \CB_0] - \CG\CB_-\CG^{-1} + \CB_+\,,
\ee
where $\CG$, viewed as the field corresponding to the gauge transformation between the two patches of $\PT_Q$, is a (degree 0) $(0,\bullet)$-form on $\PT \backslash\{Q = 0\}$ valued in the gauge group $G$.

It is useful to extract from these expressions the equations of motion of the physical fields and their gauge redundancies.
The physical (ghost number 0) fields $A_\pm$, $B_\pm$ are the 1-form parts of $\CA_\pm$, $\CB_\pm$ and the 0-form parts $g$, $B_0$ of $\CG$, $\CB_0$, respectively.
Focusing on the gauge fields, we see that the equations of motion for the physical fields $A_\pm = \CA^{(1)}_\pm$ and $g = \CG^{(0)}$ read
\be
	\ol{\pd} A_\pm + \tfrac{1}{2}[A_\pm, A_\pm] = 0 \qquad \ol{\pd} g + A_+ g - g A_- = 0
\ee
The first pair of equations say that the partial connections $\ol{\pd} + A_\pm$ define holomorphic $G$ bundles on the two copies of $\PT$. 
The latter says that $g$ is a isomorphism of these holomorphic $G$-bundles over $Q \neq 0$; acting with $g^{-1}$ on the equations of motion for $g$, we find that the partial connections are indeed gauge-equivalent away from $Q = 0$.
The infinitesimal gauge redundancies encoded in $\CQ$ are the usual ones for the connections $A_\pm$, whereas the isomorphism $g$ transforms by pre- and post-composition:
\be
	\delta A_\pm = \ol{\pd} c_\pm + [A_\pm, c_\pm] \qquad \delta g = c_+ g - g c_-
\ee
where $c_\pm = \CA^{(0)}_\pm$ is the usual BV/BRST ghost for gauge transformations.
At the level of finite transformations, the isomorphism transforms as $g \to g_+ g g_-^{-1}$ if we act with gauge transformations $g_\pm$ on the two copies of $\PT$, as expected.

We can understand the allowed singularities of the gauge field $\widehat{\CA}$ by studying the moduli space of solutions to these equations of motion on a small (formal) neighborhood of the quadric $\{Q = 0\}$.
We denote by $G(\CO_Q)$ the group of holomorphic gauge transformations defined on this neighborhood and denote by $G(\CK_Q)$ the group of holomorphic gauge transformations defined on the complement of $\{Q = 0\}$ in this neighborhood, \emph{i.e.} the (formal) punctured neighborhood of $Q = 0$.
The double quotient $G(\CO_Q) \backslash G(\CK_Q) / G(\CO_Q)$ can naturally be viewed as a bundle over the quadric $Q = 0$ and a section thereof is identified with a solution to the equations of motion on this formal neighborhood.
Each fiber of this bundle is of the form $G[[Q]] \backslash G(\!(Q)\!) / G[[Q]]$, where $G(\!(Q)\!)$ (resp. $G[[Q]]$) is the group of Laurent (resp. Taylor) series-valued group elements.
The single quotient $G(\!(Q)\!)/ G[[Q]]$ is a well-studied object in algebraic geometry known as the affine Grassmannian $\text{Gr}_G$, familiar from \emph{e.g.} the Geometric Langlands program \cite{Kapustin:2006pk}.
The fibers are glued non-trivially and a section of this bundle can be viewed as a map from the quadric to a version of the $G[[Q]]$-quotient of the affine Grassmannian twisted by the normal bundle of $\{Q = 0\}$ inside $\PT$.

The equations of motion for the physical fields in $\widehat{\CB}$ are analogous:
\be
	\ol{\pd} B_\pm + [A_\pm, B_\pm] = 0 \qquad \ol{\pd} B_0 + [A_+, B_0] + B_+ - g B_- g^{-1} = 0
\ee
Thus, the equations of motion say $B_\pm$ are adjoint-valued holomorphic sections of the holomorphic $G$-bundles defined by $\ol{\pd} + A_\pm$.
The field $B_0$ need not be holomorphic, with its failure measured by the difference of the restrictions $B_+ - g B_- g^{-1}$.
The redundancies encoded in $\CQ$ includes the gauge transformations above together with a shift symmetry:
\be
	\delta B_\pm = \ol{\pd} b_\pm + [A_\pm, b_\pm] + [c_\pm, B_\pm] \qquad \delta B_0 = [c_+, B_0] + b_+ - g b_- g^{-1}
\ee
where $b_\pm = \CB^{(0)}_\pm$.

We are ultimately interested in understanding holomorphic gauge theory perturbatively around a fixed background gauge bundle, \emph{i.e.} a fixed choice of $\CG$, say $\CG_0$.
Although not strictly necessary, we will assume $\CG_0 = g_0$ only has 0-form components so that the background only turns on physical fields.
We then write
\be
	\CG = \big(1 + \CA_0 + \dots \big) \CG_0
\ee
for a ``small'' field $\CA_0$ of ghost number 0.
The equations of motion for $\CA_0$ and redundancies thereof are then encoded in the differential
\be
	\CQ \CA_0 = \ol{\pd} \CA_0 + \tfrac{1}{2}[\CA_+ + \CA_-^{\CG_0}, \CA_0] + \CA_+ - \CA_-^{\CG_0}
\ee
where $\CA_-^{\CG_0} = \CG_0 \CA_- \CG^{-1}_0 + \CG_0 \ol{\pd} \CG_0^{-1}$ is the gauge transformation of $\CA_-$ by $\CG_0$.
More precisely, the equations of motion for $\CA_0$ arise by imposing $\CQ \CG = (\CQ \CA_0) \CG_0 + \dots$ which gives
\be
	\CQ \CA_0 = \ol{\pd} \CA_0 + \CA_+ \CA_0 - \CA_0 \CA_-^{\CG_0} + \CA_+ - \CA_-^{\CG_0}\,.
\ee
We can equivalently get the equations of motion from imposing $\CQ \CG^{-1} = - \CG_0^{-1} (\CQ \CA_0) + \dots$, which gives
\be
	\CQ \CA_0 = \ol{\pd} \CA_0 + \CA_-^{\CG_0} \CA_0 - \CA_0 \CA_+ + \CA_+ - \CA_-^{\CG_0}\,.
\ee
As these procedures should lead to the same equations of motion, we can equally take the average of these expressions, giving rise to the differential presented above.
Note, however, that the consistency of these equations also requires
\be
\label{eq:A0constraint}
	\CA_+ \CA_0 - \CA_0 \CA_-^{\CG_0} = \CA_-^{\CG_0} \CA_0 - \CA_0 \CA_+\,.
\ee

Just as above, it is useful to unpack this differential.
As $\CA_0$ is degree 0, the physical field is its zero-form part $A_0 = \CA^{(0)}_0$; the equations of motion for $A_0$ then read
\be
	\ol{\pd} A_0 + \tfrac{1}{2}[A_+ + A_-^{g_0}, A_0] + A_+ - A_-^{g_0} = 0\,.
\ee
With respect to the above gauge transformations, $A_0$ transforms as
\be
	\delta A_0 = \tfrac{1}{2}[c_+ + c_-^{g_0}, A_0] + c_+ - c_-^{g_0}
\ee
We also record the action of $\CQ$ on $\CB_0$ when working around this background:
\be
	\CQ \CB_0 = \ol{\pd} \CB_0 + [\CA_+, \CB_0] - [\CA_0, \CB_-^{\CG_0}] + \CB_+ - \CB_-^{\CG_0}
\ee

\subsection{The Coulomb bundle}
\label{sec:coulomb}
We now describe a particularly simple background field configuration that will lift the self-dual dyon to holomorphic gauge theory on $\PT_Q$.
There are many related realizations of the self-dual dyon field from twistor space, see \emph{e.g.} \cite{PS79}.
We present this complementary approach to the self-dual dyon in Appendix \ref{sec:quadrille}, following the recent paper \cite{ABMS2}.

From our analysis of holomorphic gauge theory on $\PT_Q$, we find that such a background should arise from a pair of line bundles on $\PT$ that are isomorphic over $Q \neq 0$.
Over one copy of $\PT$ we take the bundle $\CO(1) \to \PT$ and over the other we take $\CO(-1) \to \PT$.
Over $Q \neq 0$, the nowhere vanishing section $Q^{-1}$ of $\CO(-2)$ can be used to identify these bundles, \emph{i.e.} we take $g_0 = Q^{-1}$.
The resulting bundle is called the Coulomb bundle \cite{PS90} and will be denoted $\CC$.

When we include the Coulomb bundle into non-abelian gauge theory, we will simply consider $g = Q^{-T_0}$ for our preferred generator $T_0$ or, equivalently, a cocharacter of the maximal torus $T \subset G$.
Based on the description of the moduli space of solutions to the equations of motion as sections of the bundle $\CE$ over the quadric $\{Q = 0\}$, working around this background should be thought of as working in the neighborhood of a section that lies in the $G(\CO)$-orbit of $[Q^{-T_0}]$ in the affine Grassmannian $\text{Gr}_G$.
A natural extension to these line operators would be to work in other $G(\CO)$-orbits of the affine Grassmannian $\text{Gr}_G$, possibly dressed by additional $G[[Q]] \rtimes \C^\times$-equivariant%
\footnote{The additional $\C^\times$-equivariance acts on $G[[Q]]$ and $\text{Gr}_G$ by rotating $Q$ and ensures that we have the necessary symmetries to glue local expressions into something global on the quadric $Q = 0$.} %
vector bundles or sheaves, \emph{i.e.} additional Chan-Paton factors.%
\footnote{We thank Kevin Costello for bringing to our attention the appearance of the affine Grassmannian and this natural generalization of the Coulomb bundle.} %

The cohomology group $H^{(0,1)}(\PT_Q, \CC^e)$, where $\CC^e$ is the $e$th power of the Coulomb bundle, gives a twistorial realization of helicity $+1$ fields of charge $e$ in the presence of the self-dual dyon; fields of helicity $h$ can be realized by tensoring with $\CO(2h-2)$, which we denote $\CC^e(2h-2)$ for simplicity.
We can build a field-theoretic description of the cohomology groups $H^{(0,\bullet)}(\PT_Q, \CC^e)$ analogous to Section \ref{sec:sources}.
We consider fields $\CZ_\pm \in \Omega^{(0, \bullet)}(\PT, \CO(\mp e))$ defined on all of $\PT$ and a field $\CZ_0 \in \Omega^{(0, \bullet)}(\PT\backslash\{Q=0\}, \CO(-e))$ defined away from $Q = 0$.
The equations of motion are encoded in the differential
\be
	\CQ \CZ_\pm = \overline{\pd} \CZ_\pm\,, \qquad \CQ \CZ_0 = \overline{\pd}\CZ_0 + \CZ_+ - Q^{-e} \CZ_-\,.
\ee

\subsubsection{Reducing to spacetime}
\label{sec:Coulombreduction}
To see that the Coulomb bundle describes a self-dual dyon, we again use the twistor correspondence described in Section \ref{sec:sources}.
For a general abelian background $\widehat{\CA} = (\CA_\pm, \CG)$, the corresponding 4d self-dual field strength is given by the following expression:
\be
	F_{\gamma}(x) = \int_{D_+} \pd_{\dot{\alpha}} \pd_{\dot{\beta}} \CA_+ + \int_{D_-} \pd_{\dot{\alpha}} \pd_{\dot{\beta}} \CA_- + \int_\gamma \pd_{\dot{\alpha}} \pd_{\dot{\beta}} \ln \CG
\ee
Plugging in the data for the Coulomb bundle $\CA_\pm = 0$, $\CG = Q^{-1}$ we find the field strength is given by
\be
	F_{\gamma}(x) = \int_\gamma \pd_{\dot{\alpha}} \pd_{\dot{\beta}} \ln Q^{-1} = \int_\gamma \text{d} z \begin{pmatrix} 
	1 & -z\\ -z & z^2
	\end{pmatrix} Q^{-2} = 2 \pi i \begin{pmatrix}
		\frac{x^1 - i x^2}{4 r^3} & \frac{x^3}{4 r^3}\\ \frac{x^3}{4 r^3} & \frac{- x^1 - i x^2}{4 r^3}
	\end{pmatrix}
\ee
as desired.

We can similarly reduce a section of the charge $e$ Coulomb bundle $\CC^e$ to spacetime.
The main complication will be due to the fact that we must choose a trivialization of $\CC^e$ on each twistor sphere $\PP^1_{Q,x}$, but this is standard practice in twistorial realizations of self-dual gauge theories.
We cover $z \neq \ol{\zeta}^{-1}$ with two open set $V_+$, $U_+$ on which $z \neq \infty$ and $z \neq 0$; we similarly cover $z \neq \zeta$ with two open sets $V_-$, $U_-$.
The bundle on $\PP^1_{Q,x}$ we get by pulling back $\CC^e$ can be described by the following transition functions:
\be
\begin{aligned}
	c_{V_+, V_-} & = Q^{-e} & \qquad c_{V_-, U_-} & = z^{e} & \qquad c_{V_+, U_-} & = z^{e} Q^{-e}\\
	c_{U_+, U_-} & = \wt{Q}^{-e} & \qquad c_{V_+, U_+} & = z^{-e} & \qquad c_{V_-, U_+} & = z^{-e} Q^{e}\\
\end{aligned}
\ee
The first column reflects the gluing of the two copies of $\PT$ and the second column reflects the fact that the field takes values in $\CO(\mp e)$ on the two copies of $\PT$.
This bundle on $\PP^1$ is known as the twistor quadrille of Penrose and Sparling \cite{PS79}.

We can trivialize this bundle by building a nowhere vanishing section.
The construction of such a section depends on the crucial fact that $Q$ and $\wt{Q}$ can be factored away from $r = 0$:
when $\zeta \neq \infty$ we can write $Q = q_+ q_-$ where
\be
	q_+(x,z) = -\bigg(\frac{2r}{1+|\zeta|^2}\bigg)^{1/2}(1 + \ol{\zeta} z) \qquad	q_-(x,z) = \bigg(\frac{2r}{1+|\zeta|^2}\bigg)^{1/2}(z - \zeta)
\ee
and we can write $\wt{Q} = \wt{q}_+ \wt{q}_-$ where
\be
	\wt{q}_+(x,z) = -\bigg(\frac{2r}{1+|\zeta|^2}\bigg)^{1/2}(w + \ol{\zeta}) \qquad \wt{q}_-(x,z) = \bigg(\frac{2r}{1+|\zeta|^2}\bigg)^{1/2}(1 - \zeta w)\,.
\ee
We can similarly factor $Q$ and $\wt{Q}$ when $\zeta \neq 0$.

It is immediate that $q_\pm(x,z)$ doesn't vanish on $V_\pm$ and $\wt{q}_\pm(x,z)$ doesn't vanish on $U_\pm$.
Additionally, they can be expressed in terms of the pairings $\langle \lambda \chi_\pm\rangle$ of the homogeneous coordinates $\lambda_\alpha$ with spinors $\chi_\pm$
\be
	q_\pm = \frac{\langle \lambda \chi_\pm\rangle}{\lambda_2} \qquad \wt{q}_\pm = \frac{\langle \lambda \chi_\pm\rangle}{\lambda_1}
\ee
where
\be
	\chi_{+,\alpha} = \bigg(\frac{2r}{1+|\zeta|^2}\bigg)^{1/2}(-\ol{\zeta}, 1) \qquad \chi_{-,\alpha} = \bigg(\frac{2r}{1+|\zeta|^2}\bigg)^{1/2}(1, \zeta)
\ee
are the charged Killing spinors of \cite{ABMS1}.
We can build a nowhere-vanishing section of the Coulomb bundle restricted to this $\PP^1_{Q,x}$ represented locally on $V_+$, $U_+$, $V_-$, and $U_-$ as $q_+(x,z)^{-1}$, $\wt{q}_+(x,w)^{-1}$, $q_-(x,z)$, and $\wt{q}_-(x,w)$, respectively.
A nowhere-vanishing section over points with $\zeta \neq 0$ can be constructed in a similar fashion.

With this trivialization in hand, we can now extract a spacetime field from a section of the Coulomb bundle; for simplicity we focus on a field valued in $\CC^e(-2)$, corresponding to a scalar field of charge $e$.
If we denote the trivialized section by $\widehat{\CZ}^H$, and similarly for its components, we then write
\be
	Z_{\gamma}(x) = \int_{\PP^1_{Q,x}, \gamma} \widehat{\CZ}^H = \int_{D_+} \CZ_+^H + \int_{D_-} \CZ_-^H + \int_{\gamma} \CZ_0^H\,.
\ee
The resulting 4d field is gauge invariant ($\CQ$-closed) and changes by a $\CQ$-exact term when we replace $\gamma$ by a homologous curve.

\subsubsection{Quasi-momentum eigenstates}
\label{sec:quasimomentumstates}

The main states of interest are realized by taking $\CZ_\pm$ to be modified plane waves
\be
\label{eq:dressedPW}
	\CZ_+ = Q^{m} \exp(v^{\dot\alpha} \tilde{\lambda}_{\dot\alpha}) \delta(z - z_0) \,, \qquad \CZ_- = Q^{m+e} \exp(v^{\dot\alpha} \tilde{\lambda}_{\dot\alpha}) \delta(z - z_0)\,, \qquad \CZ_0 = 0\,.
\ee
For this to be a well-defined state on $\PT_Q$, both $m$ and $m + e$ must be non-negative integers.
We will call these states quasi-momentum eigenstates, in analogy with the usual momentum eigenstates, and they correspond to solutions of the massless,  background-coupled, linearized equations of motion; see Section 2.3 of \cite{ABMS2} for more details.
These will furnish the asymptotic scattering states in the presence of the self-dual dyon.

We will see the constraint $m \geq 0$ and $m + e \geq 0$ that the above fields give genuine states on $\PT_Q$ ensures the resulting 4d fields are non-singular; we will call the states saturating this bound, \emph{i.e.} satisfying $m = \max(0, -e)$, \emph{minimal} states (of charge $e$).
The non-minimal states can be obtained from the minimal ones by differentiating with respect to the parameters $\tilde{\lambda}$
\be
	Q^{m+1} \exp(v^{\dot\alpha} \tilde{\lambda}_{\dot\alpha}) = \bigg(\frac{\pd\hfill}{\pd \tilde{\lambda}_{\dot 1}} + z \frac{\pd\hfill}{\pd \tilde{\lambda}_{\dot 2}}\bigg) Q^m \exp(v^{\dot\alpha} \tilde{\lambda}_{\dot\alpha})
\ee
The above formula gives the following 4d field:
\be
	Z_{\gamma} = q_+(x,z_0)^{m + e} q_-(x,z_0)^{m} e^{i k \cdot x}
\ee
where the momentum $k = (k_{\alpha \dot \alpha})$ satisfies $k_{1 \dot \alpha} = -i \tilde{\lambda}_{\dot \alpha}$ and $k_{2 \dot \alpha} = -i z \tilde{\lambda}_{\dot \alpha}$.
These are precisely the (scalar) quasi-momentum eigenstates of \cite{ABMS1}.%
\footnote{There is a slight difference in notation that must be noted to correctly compare the two results.
Our notion of charge is required to be integral and is identified with twice their charge, which is half-integral.
Additionally, our integer parameter $m$ is identified with their $m - e$.} %
The above relation between minimal and non-minimal 6d quasi-momentum eigenstates is also realized in 4d, \emph{cf.} Eq. (3.27) of \cite{ABMS1}.
We also note that these states are single-valued and non-singular away from $r = 0$ only if $m \geq 0$ and $m + e \geq 0$ or, equivalently, $m \geq \max(0,-e)$.
Reducing the above field configuration, viewed as a section of $\CC^e(2h-2)$ for other $h$, similarly leads to the other quasi-momentum eigenstates for other helicities.

\section{Celestial Symmetries in the Presence of a Twistorial Line Defect}
\label{sec:dyonCCA}

In this section we turn to the question of how celestial symmetries of a twistorial QFT are modified in the presence of a twistorial line operator inserted along $\ell$.
As described in \cite{CP22}, there are two natural realizations of these celestial symmetries in terms of the parent 6d holomorphic field theory.
On one hand, they can be realized as gauge transformations away from points on $\PT$ with $z = 0$ or $z = \infty$ that preserve the vacuum, \emph{cf.} Section 4 of \cite{CP22}.
On the other, the full celestial chiral algebra can be realized as the algebra of local operators on a universal holomorphic defect wrapping the twistor sphere over the origin in $\R^4$ in the parent 6d holomorphic field theory on $\PT$, \emph{cf.} Section 7 of \cite{CP22}.
We propose that these two perspectives continue to hold in the presence of twistorial line operators realized from $\PT_Q$, which we illustrate with the example of the self-dual dyon.

\subsection{Celestial symmetries from gauge transformations}
\label{sec:gaugesymms}

Perhaps the most direct route to the Lie algebra of celestial symmetries of a twistorial QFT is by investigating gauge transformations on twistor space away from $z = 0$ and $z = \infty$.
The same is true for celestial symmetries in the presence of a twistorial line operator.
In the example of self-dual gauge theory around a fixed background $\CG = \CG_0 = g_0$, this (DG) Lie algebra $\mathfrak{G}$ is obtained by viewing the (graded) commutative algebra generated by $c_\pm$, $b_\pm$, $A_0$, and $B_0$ together with the differential $\CQ$ as the Chevalley-Eilenberg complex of $\mathfrak{G}$.
For ease of reading, we record those transformations again:
\be
\begin{aligned}
	\CQ c_\pm & = \tfrac{1}{2}[c_\pm, c_\pm] \qquad & \CQ A_0 & = \tfrac{1}{2}[c_+ + c_-^{g_0}, A_0] + c_+ - c_-^{g_0}\\
	\CQ b_\pm & = [c_\pm, b_\pm] \qquad & \CQ B_0 & = [c_+, B_0] - [A_0, b_-^{g_0}] + b_+ - b_-^{g_0}\\
\end{aligned}
\ee

First, there are degree 0 generators
\be
\begin{aligned}
	J^{(m)}_{a, \pm}[r_1, r_2]_k \longleftrightarrow c_\pm = z^k Q^m (v^{\dot 1})^{r_1} (v^{\dot 2})^{r_2} T_a\\
	\wt{J}^{(m)}_{a, \pm}[r_1, r_2]_k \longleftrightarrow b_\pm = z^k Q^m (v^{\dot 1})^{r_1} (v^{\dot 2})^{r_2} T_a\\
\end{aligned}
\ee
where $T_a$ are a basis for the gauge Lie algebra $\fg$ and $m, r_1, r_2 \geq 0$, $k \in \Z$.
These generators are not linearly independent, being related as
\be
	J^{(m+1)}_{a, \pm}[r_1, r_2]_k = J^{(m)}_{a, \pm}[r_1 + 1, r_2]_k + J^{(m)}_{a, \pm}[r_1, r_2 + 1]_{k+1}
\ee
and
\be
	\wt{J}^{(m+1)}_{a, \pm}[r_1, r_2]_k = \wt{J}^{(m)}_{a, \pm}[r_1 + 1, r_2]_k + \wt{J}^{(m)}_{a, \pm}[r_1, r_2 + 1]_{k+1}
\ee
because $Q = v^{\dot 1} + z v^{\dot 2}$.
The non-vanishing Lie brackets of these generators come from the quadratic terms in the action of $\CQ$ on $c_\pm$, $b_\pm$.
These are independent of the choice of background $g_0$ and take following form:
\be
\begin{aligned}
	\left[J^{(m)}_{a, \pm}[r_1, r_2]_k,  J^{(n)}_{b, \pm}[s_1, s_2]_l\right] & = f^c_{ab} J^{(m+n)}_{c, \pm}[r_1 + s_1, r_2 + s_2]_{k+l}\\
	\left[J^{(m)}_{a, \pm}[r_1, r_2]_k,  \wt{J}^{(n)}_{b, \pm}[s_1, s_2]_l\right] & = f^c_{ab} \wt{J}^{(m+n)}_{c, \pm}[r_1 + s_1, r_2 + s_2]_{k+l}
\end{aligned}
\ee
The vectors $J^{(0)}_{a, \pm}[r_1, r_2]_k$ and $\wt{J}^{(0)}_{a, \pm}[r_1, r_2]_k$ suffice to span the degree 0 generators; they are the ordinary conformally soft symmetries of pure, self-dual gauge theory found in \cite{Guevara:2021abz} (see also \cite{Adamo:2021zpw}).
Another convenient basis of this degree 0 subspace is with the generators $J^{(m)}_{a, \pm}[0, r]_k$ and $\wt{J}^{(m)}_{a, \pm}[0, r]_k$.

There are additionally degree $-1$ generators coming from $A_0$ and $B_0$:
\be
\begin{aligned}
	\psi^{(m)}_{a, \pm}[r_1, r_2]_k \longleftrightarrow A_0 = z^k Q^m (v^{\dot 1})^{r_1} (v^{\dot 2})^{r_2} T_a\\
	\wt{\psi}^{(m)}_{a, \pm}[r_1, r_2]_k \longleftrightarrow B_0 = z^k Q^m (v^{\dot 1})^{r_1} (v^{\dot 2})^{r_2} T_a
\end{aligned}
\ee
where $r_1, r_2 \geq 0$ and $m, k \in \Z$.
Again, these generators are not linearly independent and are related as above; note, however, that it does not suffice to consider the $m = 0$ states as $m$ can be any integer.
Instead, we can realize a basis of this degree $-1$ subspace with the generators $\psi^{(m)}_a[0,r]_k$ and $\wt{\psi}^{(m)}_a[0,r]_k$.
These generators will always bracket trivially with one another -- there are no generators in degrees other than $0$ and $-1$.
The non-vanishing brackets with the degree $0$ generators are governed the quadratic terms in the action of $\CQ$ on $A_0$ and $B_0$ and generally depend on the background $g_0$.
For the self-dual dyon $g_0 = Q^{-T_0}$, these brackets are given as follows:
\be
\begin{aligned}
	\left[J^{(m)}_{a, +}[r_1, r_2]_k,  \psi^{(n)}_{b}[s_1, s_2]_l\right] & = \tfrac{1}{2} f^c_{ab} \psi^{(m+n)}_{c}[r_1 + s_1, r_2 + s_2]_{k+l}\\
	\left[J^{(m)}_{a, -}[r_1, r_2]_k,  \psi^{(n)}_{b}[s_1, s_2]_l\right] & = \tfrac{1}{2} f^c_{ab} \psi^{(m+n-e_a)}_{c}[r_1 + s_1, r_2 + s_2]_{k+l}\\
	\left[\wt{J}^{(m)}_{a, -}[r_1, r_2]_k,  \psi^{(n)}_{b}[s_1, s_2]_l\right] & = f^c_{ab} \wt{\psi}^{(m+n-e_a)}_{c, \pm}[r_1 + s_1, r_2 + s_2]_{k+l}\\
	\left[J^{(m)}_{a, +}[r_1, r_2]_k,  \wt{\psi}^{(n)}_{b}[s_1, s_2]_l\right] & = f^c_{ab} \wt{\psi}^{(m+n)}_{c, \pm}[r_1 + s_1, r_2 + s_2]_{k+l}\\
\end{aligned}
\ee
where $f^c_{ab}$ are the structure constants of $\fg$ in the basis $T_a$.
The factors of $\tfrac{1}{2}$ appearing in the action of $J^{(m)}_{a, \pm}[r_1, r_2]_k$ on $\psi^{(n)}_b[s_1, s_2]$ might give pause, but this it is a non-issue in this case because the left and right actions of gauge transformations on $\CA_0$ must agree, \emph{cf.} Eq. \eqref{eq:A0constraint} --- the factors of $\tfrac{1}{2}f^c_{ab}$ can equivalently be replaced by either of these actions.

Finally, there is a background-dependent differential acting on these generators determined by the linear terms in the action of $\CQ$.
By an abuse of notation, we also call this differential $\CQ$.
For the self-dual dyon, the action of $\CQ$ on the above generators is given by
\be
\begin{aligned}
	\CQ J^{(m)}_{a, \pm}[0, r]_k & = 0\\
	\CQ \psi^{(m)}_a[0, r] & = \begin{cases}
		J^{(m)}_{a, +}[0, r]_k - J^{(m + e_a)}_{a, -}[0, r]_k  & m \,, m + e_a \geq 0\\
		- J^{(m + e_a)}_{a, -}[0, r]_k  & m < 0 \,, m + e_a \geq 0\\
		J^{(m)}_{a, +}[0, r]_k & m \geq 0 \,, m + e_a < 0\\
		0 & m \,, m + e_a < 0
	\end{cases}\\
	\CQ \wt{J}^{(m)}_{a, \pm}[0, r]_k & = 0\\
	\CQ \wt{\psi}^{(m)}_a[0, r] & = \begin{cases}
		\wt{J}^{(m)}_{a, +}[0, r]_k - \wt{J}^{(m + e_a)}_{a, -}[0, r]_k  & m \,, m + e_a \geq 0\\
		- \wt{J}^{(m + e_a)}_{a, -}[0, r]_k  & m < 0 \,, m + e_a \geq 0\\
		\wt{J}^{(m)}_{a, +}[0, r]_k & m \geq 0 \,, m + e_a < 0\\
		0 & m \,, m + e_a < 0
	\end{cases}
\end{aligned}
\ee

We will mostly be interested in the degree zero cohomology of this DG Lie algebra, which is itself a Lie algebra and what most might call the algebra of celestial symmetries.
For the self-dual dyon, this Lie algebra is spanned by \emph{e.g.} the operators $J^{\text{min}}_a[r_1, r_2]_k = J^{(m_a)}_{a, +}[r_1, r_2]_k$ with $m_a = \max(0, -e_a)$.
We will give an explicit presentation of the Lie brackets of these generators as OPEs of currents in the celestial chiral algebra in at the end of this section.
Note that this Lie algebra can be identified with a subalgebra of the full Lie algebra of celestial symmetries of self-dual gauge theory.
This is a generic feature: the (degree 0 cohomology of the) algebra of celestial symmetries in the presence of a twistorial line defects can be identified with a (defect-dependent) subalgebra of the algebra of celestial symmetries in the absence of line defects.

\subsection{Celestial symmetries from Koszul duality}
\label{sec:KDsymms}

Our second method for accessing the algebra of celestial symmetries is Koszul duality.
Namely, the algebra of celestial symmetries of a 4d twistorial QFT can be organized into a celestial chiral algebra which itself can be identified with the algebra of local operators on a universal 2d holomorphic defect inserted on a twistor sphere in the parent 6d holomorphic QFT on twistor space.
In brief, the algebra of local operators on this universal defect is the most general algebra that can be coupled to bulk local operators restricted to the location of the defect.
As described in \cite{GO19, CP20}, the notion of Koszul duality neatly encodes these constraints: the algebra of local operators on the universal defect is the Koszul dual of the algebra of bulk local operators restricted to the location of the defect.
See \emph{e.g.} \cite{PW21} for a more detailed exposition of this connection.
We again illustrate this method with our running example of the self-dual dyon.

\subsubsection{A Koszul duality refresher}

Before analyzing the universal 2d holomorphic defect of holomorphic gauge theory placed on twistor spheres in $\PT_Q$, we first recall the case of twistor spheres in $\PT$ described in \cite{CP22}.
When we place the universal 2d holomorphic defect on a twistor sphere $\PP^1_x$, there are operators $J_a[r_1, r_2]$ and $\wt{J}_a[r_1, r_2]$ that couple to the bulk fields $\CA^a$ and $\CB^a$ via the interaction
\be
	\sum_{r_1, r_2 \geq 0} \int_{\PP^1_x} J_a[r_1, r_2] D_r \CA^a + \wt{J}_a[r_1, r_2] D_r \CB^a
\ee
where $D_r$, with $r = (r_1, r_2)$, is shorthand for the following differential operator:
\be
	D_r = \frac{1}{r_1! r_2!} \bigg(\frac{\pd \hfill}{\pd v^{\dot 1}}\bigg)^{r_1}\bigg(\frac{\pd \hfill}{\pd v^{\dot 2}}\bigg)^{r_2}
\ee
To ensure that this coupling is non-vanishing, \emph{i.e.} the integrand can be identified with a $(1, \bullet)$-form, the operator $J_a[r_1, r_2]$ (resp. $\wt{J}_a[r_1, r_2]$) must transform as a section of $\CO(-2 + r_1 + r_2)$ (resp. $\CO(2 + r_1 + r_2)$) on $\PP^1_x$.

The most general holomorphic defect that holomorphic gauge theory can be coupled to simply requires that the above interaction is gauge invariant, with no other constraints on the operators $J_a[r_1, r_2]$ and $\wt{J}_a[r_1, r_2]$.
For simplicity, let's focus on the operators that couple to $\CA$ and work at tree level.
Gauge-invariance of this interaction requires that the (path-ordered) exponential of the interaction is $\CQ$-closed, which constrains the OPEs of the operators $J_a[r_1, r_2]$ with one another.
Explicitly, these constraints arise as follows:
\be
\begin{aligned}
	0 & = \CQ\bigg(1 + \sum_{r_1, r_2} \int_{w} J_a[r_1, r_2](w) D_r \CA^a(x,w)\\
	& \qquad + \tfrac{1}{2}\sum_{\substack{r_1, r_2\\ s_1, s_2\\}} \int_{z \neq w} J_a[r_1, r_2](z) J_b[s_1, s_2](w)  D_r \CA^a(x,z) D_s \CA^b(x,w) + \dots\bigg)\\
	& = - \sum_{r_1, r_2} \int_{z} J_a[r_1, r_2](w) D_r \big(\dots + \tfrac{1}{2} f^a_{bc} \CA^b(x,w) \CA^c(x,w)\big)\\
	& \qquad + \sum_{\substack{r_1, r_2\\ s_1, s_2\\}} \int_{z \neq w} J_a[r_1, r_2](z) J_b[s_1, s_2](w)  D_r \big(\ol{\pd} \CA^a(x,z) + \dots\big) D_s \CA^b(x,w) + \dots\\
	& = -\tfrac{1}{2}\sum_{\substack{r_1, r_2\\ s_1, s_2\\}} \int_w f^a_{bc} J_a[r_1 + s_1, r_2 + s_2](w) D_r \CA^b(x,w) D_s \CA^c(x,w)\\
	& \qquad + \tfrac{1}{2}\sum_{\substack{r_1, r_2\\ s_1, s_2\\}} \int_w \oint_{z = w} J_a[r_1, r_2](z) J_b[s_1, s_2](w) D_r \CA^a(x,w) D_s \CA^b(x,w) + \dots
\end{aligned}
\ee
As this must be true for all values of $\CA$, we see that gauge-invariance imposes the following OPEs (absorbing factors of $2 \pi i$ into the definition of the operators):
\be
	J_a[r_1, r_2](z) J_b[s_1, s_2](w) \sim \frac{f^c_{ab}}{z-w} J_c[r_1 + s_1, r_2 + s_2](w)
\ee
The remaining OPEs are determined in a similar fashion: the OPE of $J_a$ and $\wt{J}_b$ takes the form
\be
	J_a[r_1, r_2](z) \wt{J}_b[s_1, s_2](w) \sim \frac{f^c_{ab}}{z-w} \wt{J}_c[r_1 + s_1, r_2 + s_2](w)
\ee
and the OPE of $\wt{J}_a$ and $\wt{J}_b$ is regular.

We will often be interested in the operators sourced by the states with $\CA = \exp(\tilde{\lambda}\cdot v) \delta(z-z_0) T_a$ and with $\CB = \exp(\tilde{\lambda} \cdot v) \delta(z - z_0)$; these correspond to 4d momentum eigenstates under the twistor correspondence.
The positive-helicity plane waves source the operators
\be
	J_a[\tilde{\lambda}](z_0) = \exp(i k \cdot x) \sum_{r_1, r_2} \frac{(\tilde{\lambda}_{\dot 1})^{r_1} (\tilde{\lambda}_{\dot 2})^{r_2}}{r_1! r_2!} J_a[r_1, r_2](z_0)\,.
\ee
The exponential factor shows up from the restriction of $\exp(\tilde{\lambda} \cdot v)$ (and its derivatives) to the point $z_0$ on the twistor sphere $\PP^1_x$.
Similarly, the negative-helicity plane waves source the operators
\be
	\wt{J}_a[\tilde{\lambda}](z_0) = \exp(i k \cdot x) \sum_{r_1, r_2} \frac{(\tilde{\lambda}_{\dot 1})^{r_1} (\tilde{\lambda}_{\dot 2})^{r_2}}{r_1! r_2!} \wt{J}_a[r_1, r_2](z_0)\,.
\ee
The OPE of two positive-helicity plane-wave generators takes the form
\be
	J_a[\tilde{\lambda}](z) J_b[\tilde{\lambda}'](w) \sim \frac{f^c_{ab}}{z-w} J_c[\tilde{\lambda}+\tilde{\lambda}'](w)
\ee
The OPE plane-wave generators with opposite helicities is given by
\be
	J_a[\tilde{\lambda}](z) \wt{J}_b[\tilde{\lambda}'](w) \sim \frac{f^c_{ab}}{z-w} \wt{J}_c[\tilde{\lambda}+\tilde{\lambda}'](w)
\ee
and the OPE of two negative-helicity plane-wave generators is regular.

\subsubsection{The universal holomorphic defect on \texorpdfstring{$\PP^1_{Q,x}$}{P1Qx}}

We can repeat the above analysis on twistor spheres in $\PT_Q$ to account for the presence of sources.
The main difference will be kinematical; for simplicity, consider the case of a free scalar field realized on $\PT_Q$ by the field $\widehat{\CZ}$. 
The most general coupling to operators living on $\PP^1_{Q,x}$ takes the form
\be
\begin{aligned}
	\CI & = \sum_{r_1, r_2 \geq 0} \bigg(\int_{D_+} J_+[r_1, r_2] D_{r} \CZ_+ + \int_{D_-} J_-[r_1, r_2] D_{r} \CZ_-\\
	& \qquad + \sum_{m \in \Z} \int_{\gamma} \psi^{(m)}_+[r_1, r_2] D_r (Q^{-m} \CZ_+) + \psi^{(m)}_-[r_1, r_2] D_r (Q^{-m} \CZ_-) + J^{(m)}_0[r_1, r_2] D_r (Q^{-m} \CZ_0)\bigg)
\end{aligned}
\ee
We are free to include the factors of $Q^{-m}$ for $m \in \Z$ because the curve $\gamma$ is always contained in the subspace with $Q \neq 0$.
To ensure this coupling has ghost number 0, we see that the operators $J_\pm[r_1, r_2]$, $J^{(m)}_0[r_1, r_2]$ must also be degree 0 whereas $\psi^{(m)}_\pm[r_1, r_2]$ must be degree $-1$.
The bulk fields appearing in the coupling are redundant, and so we should identify the corresponding operators.
This identification takes the form
\be
\begin{aligned}
	\psi^{(m+1)}_\pm[r_1, r_2](z) = \psi^{(m)}_\pm[r_1 + 1, r_2](z) + z \psi^{(m)}_\pm[r_1, r_2 + 1](z)\\
	J^{(m+1)}_0[r_1, r_2](z) = J^{(m)}_0[r_1 + 1, r_2](z) + z J^{(m)}_0[r_1, r_2 + 1](z)
\end{aligned}
\ee
We also introduce the notation $J^{(m)}_{\pm}[r_1, r_2]$ for $m \geq 0$ by the same recurrence relations, where $J^{(0)}_\pm[r_1, r_2] := J_\pm[r_1, r_2]$; a convenient set of generators is thus given by the degree $0$ generators $J^{(m)}_{\pm}[0, r]$ and $J^{(m)}_{0}[0, r]$ together with the degree $-1$ generators $\psi_\pm^{(m)}[0, r]$.

This coupling is highly constrained by requiring invariance under gauge transformations, \emph{i.e.} that it is $\CQ$-closed.
We will assume that the defect operators $J_\pm[r_1, r_2]$, $J_0^{(m)}[r_1, r_2]$, and $\psi^{(m)}_\pm[r_1, r_2]$ are holomorphic on the twistor sphere $\PP^1_z$.
As we will see, it is important to allow for a non-trivial internal differential acting on the currents $J_\pm$, $J_0$, $\psi_\pm$.
We compute:
\be
\begin{aligned}
	\CQ \CI & = - \sum_{r_1, r_2 \geq 0} \int_{D_+} (\CQ J_+[r_1, r_2]) D_r \CZ_+ + \int_{D_-} (\CQ J_-[r_1, r_2]) D_r \CZ_+\\
	&  \qquad + \sum_{r_1, r_2 \geq 0} \sum_{m \in \Z} \int_{\gamma} (\CQ J^{(m)}_0[r_1, r_2]) D_r (Q^{-m} \CZ_0)\\
	& \qquad + \sum_{r_1, r_2 \geq 0} - \int_\gamma J_+[r_1, r_2] D_r \CZ_+ + \sum_{m \in \Z} \int_{\gamma} \left(\CQ \psi^{(m)}_+[r_1, r_2] + J^{(m)}_0[r_1, r_2]\right) D_r (Q^{-m} \CZ_+)\\
	& \qquad + \sum_{r_1, r_2 \geq 0} \int_\gamma J_-[r_1, r_2] D_r \CZ_- + \sum_{m \in \Z} \int_{\gamma} \left(\CQ \psi^{(m)}_-[r_1, r_2] - J^{(m)}_0[r_1, r_2]\right) D_r (Q^{-m} \CZ_-)\\
\end{aligned}
\ee
Gauge invariance of this coupling dictates that the operators $J_\pm[r_1, r_2]$, $J_0[r_1, r_2]$ are $\CQ$-closed whereas $\psi_\pm[r_1, r_2]$ are generally not:
\be
\begin{aligned}
	\CQ J^{(m)}_{\pm}[0, r] = 0 \qquad \CQ J^{(m)}_{0}[0, r] = 0 \hspace{1.5cm}\\
	\CQ \psi^{(m)}_\pm[0, r] = \begin{cases}
		\pm \left(J_\pm^{(m)}[0, r] - J_0^{(m)}[0, r]\right) & m \geq 0\\
		\mp J^{(m)}_0[0, r] & m < 0\\ 
	\end{cases}
\end{aligned}
\ee
The generators $\tfrac{1}{2}(\psi^{(m)}_+[0, r] - \psi^{(m)}_-[0, r])$ trivialize the operators $\tfrac{1}{2}(J^{(m)}_+[0, r] + J^{(m)}_-[0, r]) - J^{(m)}_0[0,r]$ with $m \geq 0$ and the operators $J^{(m)}_0[0,r]$ otherwise; we see that it suffices to consider the operators $\psi^{(m)}[0,r] = \psi^{(m)}_+[0,r] + \psi^{(m)}_-[0,r]$ together with the $J^{(m)}_\pm[0,r]$.

The situation is more interesting when we couple the scalar to charge $e$ background Coulomb bundle.
Starting from the same coupling, we find that it is gauge invariant if the differential acts as
\be
\begin{aligned}
	\CQ J^{(m)}_{\pm}[0, r] = 0 \qquad \CQ J^{(m)}_{0}[0, r] = 0 \hspace{1.5cm}\\
	\CQ \psi^{(m)}_+[0, r] = \begin{cases}
		\left(J_+^{(m)}[0, r] - J_0^{(m)}[0, r]\right) & m \geq 0\\
		-J^{(m)}_0[0, r] & m < 0\\ 
	\end{cases}\\
	\CQ \psi^{(m)}_-[0, r] = \begin{cases}
		-\left(J_-^{(m)}[0, r] - J_0^{(m-e)}[0, r]\right) & m \geq 0\\
		J^{(m-e)}_0[0, r] & m < 0\\ 
	\end{cases}
\end{aligned}
\ee
In particular, we find that $\tfrac{1}{2}(\psi^{(m)}_+[0,r] + \psi^{(m+e)}_-[0,r])$ trivializes $\tfrac{1}{2}(J^{(m)}_+[0, r] + J^{(m+e)}_-[0, r]) - J^{(m)}_0[0,r]$ with $m \geq 0$ and the operators $J^{(m)}_0[0,r]$ otherwise.
It thus suffices to consider the operators $\psi^{(m)}[0,r] = \psi^{(m)}_+[0,r] + \psi^{(m+e)}_-[0,r]$ together with $J^{(m)}_\pm[0,r]$, subject to the differential
\be
\begin{aligned}
	\CQ J^{(m)}_{\pm}[0, r] & = 0\\
	\CQ \psi^{(m)}[0, r] & = \begin{cases}
		\left(J_+^{(m)}[0, r] - J_-^{(m+e)}[0, r]\right) & m\,, m+e \geq 0\\
		J_+^{(m)}[0, r] & m \geq 0 \,, m+e < 0\\
		- J_-^{(m+e)}[0, r] & m < 0\,, m+e \geq 0\\
		0 & m, m+e < 0\\
	\end{cases}
\end{aligned}
\ee
This reduced collection of operators couples to the bulk fields as follows:
\be
\begin{aligned}
	\CI = \sum_{r_1, r_2 \geq 0} & \bigg(\int_{D_+} J_+[r_1, r_2] D_{r} \CZ_+ + \int_{D_-} J_-[r_1, r_2] D_{r} \CZ_-\\
	& + \tfrac{1}{2} \int_{\gamma} J_+[r_1, r_2] D_r \CZ_0 + J_-[r_1, r_2] D_r (Q^{e} \CZ_0)\\
	& + \tfrac{1}{2} \sum_{m \in \Z} \int_{\gamma} \psi^{(m)}[r_1, r_2] D_r (Q^{-m} \CZ_+ + Q^{-m-e} \CZ_-)\bigg)
\end{aligned}
\ee

We now turn to holomorphic gauge theory.
The above kinematics persist to this case; we get degree 0 defect operators $J^{(m)}_{a, \pm}[r_1, r_2]$ and $\wt{J}^{(m)}_{a,\pm}[r_1, r_2]$ as well as degree $-1$ defect operators $\psi^{(m)}_a[r_1, r_2]$ and $\wt{\psi}^{(m)}_a[r_1, r_2]$ subject to the usual redundancies; it suffices to consider \emph{e.g.} the operators with $r_1 = 0$.
Additionally, there is a non-trivial differential $\CQ$ acting as above.
The OPEs of these generators are deduced as in the refresher above, so we will not show the details of the computation.
First, the operators $J_{a, \pm}[r_1, r_2]$ and $\wt{J}_{a, \pm}[r_1, r_2]$ realize two copies of the ordinary celestial chiral algebra for gauge theory; this is because their couplings to $\CA_{a, \pm}$ and $\CB_{a, \pm}$ are unchanged, as are the gauge transformations of these fields.
The OPEs of the degree $0$ operators with the degree $-1$ operators are ultimately dictated by the form of gauge transformations on the fields $\CA_0$ and $\CB_0$.
These OPEs are given by
\be
\begin{aligned}
	J^{(m)}_{a, +}[r_1, r_2](z) \psi^{(n)}_{b}[s_1, s_2](w) & \sim \frac{\tfrac{1}{2} f^c_{ab}}{z-w} \psi^{(m+n)}_c[r_1 + s_1, r_2 + s_2](w)\\
	J^{(m)}_{a, -}[r_1, r_2](z) \psi^{(n)}_{b}[s_1, s_2](w) & \sim \frac{\tfrac{1}{2} f^c_{ab}}{z-w} \psi^{(m+n-e_a)}_c[r_1 + s_1, r_2 + s_2](w)\\
	\wt{J}^{(m)}_{a, -}[r_1, r_2](z) \psi^{(n)}_{b}[s_1, s_2](w) & \sim \frac{f^c_{ab}}{z-w} \wt{\psi}^{(m+n-e_a)}_c[r_1 + s_1, r_2 + s_2](w)\\
	J^{(m)}_{a, +}[r_1, r_2](z) \wt{\psi}^{(n)}_{b}[s_1, s_2](w) & \sim \frac{f^c_{ab}}{z-w} \wt{\psi}^{(m+n)}_c[r_1 + s_1, r_2 + s_2](w)\\
\end{aligned}
\ee
with the remaining OPEs being regular.
Finally, the OPEs of the degree $-1$ operators $\psi^{(m)}_a[r_1, r_2]$ with one another are all regular.
Perhaps unsurprisingly, the (DG) Lie algebra of modes of this (DG) chiral algebra is precisely the same as the Lie algebra found in Section \ref{sec:gaugesymms}.

\subsection{Conformally soft generators}

As we have seen, celestial symmetries in the presence of a twistorial line operator are naturally graded by ghost number and are equipped with a differential (of ghost number $1$).
In most circumstances, the main symmetries of interest are the degree 0 cohomology of this DG celestial chiral algebra.
For the example of the celestial chiral algebra around the self-dual dyon, the differential trivializes all $J^{(m)}_{a, +}[r_1, r_2]$ and $\wt{J}^{(m)}_{a, +}[r_1, r_2]$ with $m < -e_a$, which is only interesting if $e_a$ is negative.
Similarly, it makes all $J^{(m)}_{a, -}[r_1, r_2]$ and $\wt{J}^{(m)}_{a, -}[r_1, r_2]$ with $m < e_a$ $\CQ$-exact, which can only happen if $e_a$ is positive.
The operators $J^{(m)}_{a, +}[r_1, r_2]$ and $J^{(m+e_a)}_{a, -}[r_1, r_2]$ are cohomologous so long as $m$ and $m + e_a$ are non-negative, with the same being true for $\wt{J}^{(m)}_{a, +}[r_1, r_2]$ and $\wt{J}^{(m+e_a)}_{a, -}[r_1, r_2]$.
The degree 0 cohomology is thus generated by \emph{e.g.} the operators $J^{\text{min}}_a[r_1, r_2] = J^{(m_a)}_{a, +}[r_1, r_2] = J^{(m_a + e_a)}_{a, -}[r_1, r_2] + \CQ(...)$ and the operators $\wt{J}^{\text{min}}_a[r_1, r_2] = \wt{J}^{(m_a)}_{a, +}[r_1, r_2] = \wt{J}^{(m_a + e_a)}_{a, -}[r_1, r_2] + \CQ(...)$ where $m_a = \max(0, -e_a)$.
The symmetries generated by these currents are analogous to ordinary conformally soft symmetries of self-dual gauge theory.

The OPEs of the conformally soft currents coming from positive-helicity gluons take the form
\be
\begin{aligned}
	& J^{\text{min}}_a[r_1, r_2](z) J^{\text{min}}_b[s_1, s_2](w)\\
	& \hspace{1cm} \sim \begin{cases}
		{\displaystyle \frac{f^c_{ab}}{z-w} \sum\limits_{l=0}^{m_{ab}} \binom{m_{ab}}{l} w^l J^{\text{min}}_c[r_1 + s_1 + m_{ab} - l, r_2 + s_2 + l](w)} & e_a e_b < 0\\
		{\displaystyle \frac{f^c_{ab}}{z-w} J^{\text{min}}_c[r_1 + s_1, r_2 + s_2](w)} & e_a e_b \geq 0
	\end{cases}
\end{aligned}
\ee
where $m_{ab} = \min(|e_a|, |e_b|)$.
The conformally soft currents coming from negative helicity gluons are nearly identical.
We will denote them by $\wt{J}^{\text{min}}_a[r_1, r_2](z)$.
Their OPE with the positive helicity currents takes the form
\be
\begin{aligned}
	& J^{\text{min}}_a[r_1, r_2](z) \wt{J}^{\text{min}}_b[s_1, s_2](w)\\
	& \hspace{1cm} \sim \begin{cases}
		{\displaystyle \frac{f^c_{ab}}{z-w} \sum\limits_{l=0}^{m_{ab}} \binom{m_{ab}}{l} w^l \wt{J}^{\text{min}}_c[r_1 + s_1 + m_{ab} - l, r_2 + s_2 + l](w)} & e_a e_b < 0\\
		{\displaystyle \frac{f^c_{ab}}{z-w} \wt{J}^{\text{min}}_c[r_1 + s_1, r_2 + s_2](w)} & e_a e_b \geq 0
	\end{cases}
\end{aligned}
\ee
and their OPE with one another is regular
\be
	\wt{J}^{\text{min}}_a[r_1, r_2](z) \wt{J}^{\text{min}}_b[s_1, s_2](w) \sim 0\,.
\ee

We end this section by noting that it is often convenient to resum the conformally soft generators into analogues of momentum eigenstates as
\be
	J^{\text{min}}_a[\tilde{\lambda}](z) = \exp(i k \cdot x) \sum_{r_1, r_2} \frac{(\tilde{\lambda}_{\dot 1})^{r_1} (\tilde{\lambda}_{\dot 2})^{r_2}}{r_1! r_2!} J^{\text{min}}_a[r_1, r_2](z)
\ee
and
\be
	\wt{J}^{\text{min}}_a[\tilde{\lambda}](z) = \exp(i k \cdot x) \sum_{r_1, r_2} \frac{(\tilde{\lambda}_{\dot 1})^{r_1} (\tilde{\lambda}_{\dot 2})^{r_2}}{r_1! r_2!} \wt{J}^{\text{min}}_a[r_1, r_2](z)\,.
\ee
Indeed, these are the operators sourced by the \emph{minimal} quasi-momentum eigenstates described in Section \ref{sec:quasimomentumstates}.
The non-minimal states source operators $J^{(m)}_a[\tilde{\lambda}](z)$ with $m \geq m_a$, where $J^{\text{min}}_a[\tilde{\lambda}](z) = J^{(m_a)}_a[\tilde{\lambda}](z)$, defined by translating the above recurrence relation:
\be
	J^{(m+1)}_a[\tilde{\lambda}](z) = \bigg(\frac{\pd\hfill}{\pd \tilde{\lambda}_{\dot 1}} + z\frac{\pd\hfill}{\pd \tilde{\lambda}_{\dot 2}}\bigg) J^{(m)}_a[\tilde{\lambda}](z)
\ee
This exactly mirrors the relation between minimal and non-minimal quasi-momentum eigenstates established in Section \ref{sec:quasimomentumstates}.
The same is true for the negative-helicity states, which give rise to operators $\wt{J}^{(m)}_a[\tilde{\lambda}](z)$ with $m \geq m_a$.

The OPE of two operators dual to positive-helicity quasi-momentum eigenstates is given by
\be
\begin{aligned}
	J^{(m)}_{a}[\tilde{\lambda}](z) J^{(n)}_{b}[\tilde{\lambda}'](w) & \sim \frac{1}{z-w}f^c_{ab}J^{(m+n)}_c[\tilde{\lambda}+\tilde{\lambda}'](w)\,,
\end{aligned}
\ee
in agreement with the splitting functions found in Eq. (4.7) of \cite{ABMS2}.
The OPE of two states with opposite helicity takes the form
\be
	J^{(m)}_{a}[\tilde{\lambda}](z) \wt{J}^{(n)}_{b}[\tilde{\lambda}'](w) \sim \frac{1}{z-w}f^c_{a b} \wt{J}^{(m+n)}_c[\tilde{\lambda}+\tilde{\lambda}'](w)\,,
\ee
agreeing with the splitting functions in Eq. (4.9) of \cite{ABMS2}.
Finally, the OPE of two operators dual to negative-helicity quasi-momentum eigenstates is regular
\be
	\wt{J}^{(m)}_{a}[\tilde{\lambda}](z) \wt{J}^{(n)}_{b}[\tilde{\lambda}'](w) \sim 0\,.
\ee

\section{Tree-level MHV Scattering off of a Self-dual Dyon}

In this final section we consider an application of the tools we have developed to describe scattering in the presence of twistorial line operators.
Namely, we use them to describe tree-level MHV scattering amplitudes of gluons in the background of a self-dual dyon.
To do so, we realize Yang-Mills theory (with a certain $\theta$-angle) by deforming away from the self-dual theory by the operator $\Tr(B^2)$; for tree-level MHV scattering, it suffices to consider scattering in the self-dual theory with an insertion of the operator $\Tr(B^2)$.
Tree-level MHV scattering amplitudes can then be obtained by integrating these form factors over the insertion point of this local operator.
We can arrive at these amplitudes in two ways.
First, we realize these amplitudes by way of the chiral algebra bootstrap following \cite{CP22}.
Namely, we view these form factors as correlation functions for our celestial chiral algebra in a suitable conformal block from which we show that the $n$-point MHV amplitudes of \cite{ABMS2} are determined (up to a numerical prefactor) by spacetime and celestial symmetries.
We then repeat the analysis of \cite{BC} in the presence of the Coulomb bundle and extract the $n$-point amplitudes directly from the lift of $\Tr(B^2)$ to $\PT_Q$.

\subsection{Rotating and scaling the quadric}
Recall that twistor space $\PT$ has an $SL(2,\C) \times SL(2,\C) \times \C^\times$ symmetry realizing (complexified) Lorentz transformations and scaling of spacetime.
The first $SL(2,\C)$ factor rotates the $\PP^1$ base of $\PT$ in the standard fashion and gives the $v^{\dot{\alpha}}$ spin $\frac{1}{2}$; it is generated by the holomorphic vector fields
\be
	L_k = -z^{k+1} \pd_z - \frac{k+1}{2} z^k v^{\dot{\alpha}} \pd_{\dot{\alpha}}\,.
\ee
for $k = -1, 0, 1$.
The second $SL(2,\C)$ factor rotates the fiber coordinates as a doublet, preserving the $\PP^1$ base; it is generated by the holomorphic vector fields
\be
	\widehat{L}_0 = -\tfrac{1}{2}\big(v^{\dot{1}} \pd_{\dot{1}} - v^{\dot{2}} \pd_{\dot{2}}\big) \qquad \widehat{L}_{-1} = v^{\dot{2}} \pd_{\dot{1}} \qquad \widehat{L}_{1} = -v^{\dot{1}} \pd_{\dot{2}}
\ee
Finally, the copy of $\C^\times$ preserves the $\PP^1$ base and scales the fiber coordinates with weight $1$. 

Unsurprisingly, neither of the factors of $SL(2,\C)$ preserves the quadric $\{Q = 0\}$ -- this is because both of these copies of $SL(2,\C)$ rotate the $x^4$ axis.
Only the diagonal $SL(2,\C)$ generated by $L^{\text{tot}}_k = L_k + \widehat{L}_k$ is preserved.
For example, the vector field generating the diagonal torus
\be
	L^{\text{tot}}_0 = L_0 + \widehat{L}_0 = -z \pd_z - v^{\dot{1}} \pd_{\dot{1}}
\ee
gives $v^{\dot{1}}$ spin $1$ and $v^{\dot{2}}$ spin $0$.

The $SL(2,\C) \times \C^\times$ preserving the quadric will act on the chiral algebra found above -- the factor of $SL(2,\C)$ rotates the chiral algebra plane and the factor of $\C^\times$ acts as a flavor symmetry.
(Expressing the Virasoro generators $L_{k}, \  k \geq 2$ as holomorphic vector fields on twistor space shows that they do not preserve $Q=0$, so we cannot extend this action to the full 2d conformal algebra.)

In the coordinates $(x,z)$, rotations preserving the worldline $\ell$, \emph{i.e.} the compact part of the above $SL(2,\C)$, are generated by the vector fields
\be
\begin{aligned}
	V_3 & = i (z \pd_z - \ol{z} \pd_{\ol{z}}) + x^1 \pd_{x^2} - x^2 \pd_{x^2}\,,\\
	V_2 & = -\tfrac{1}{2}(1+z^2) \pd_z -\tfrac{1}{2}(1+\ol{z}^2) \pd_{\ol z}  + x^3 \pd_{x^1} - x^1 \pd_{x^3}\,,\\
	V_1 & = \tfrac{i}{2}(1-z^2) \pd_z - \tfrac{i}{2}(1-\ol{z}^2) \pd_{\ol z} + x^2 \pd_{x^3} - x^3 \pd_{x^2}\,.\\
\end{aligned}
\ee
If we instead replace $x$ by the coordinates $x^4$, $r$, $\zeta$, $\ol{\zeta}$, they take a particularly simple form:
\be
\begin{aligned}
	V_3 & = i (z \pd_z - \ol{z} \pd_{\ol{z}}) + i (\zeta \pd_\zeta - \ol{\zeta} \pd_{\ol{\zeta}})\\
	V_2 & = -\tfrac{1}{2}(1+z^2) \pd_z -\tfrac{1}{2}(1+\ol{z}^2) \pd_{\ol z}  -\tfrac{1}{2}(1+\zeta^2) \pd_\zeta -\tfrac{1}{2}(1+\ol{z}^2) \pd_{\ol \zeta}\\
	V_1 & = \tfrac{i}{2}(1-\zeta^2) \pd_z - \tfrac{i}{2}(1-\ol{\zeta}^2) \pd_{\ol \zeta} + \tfrac{i}{2}(1-\zeta^2) \pd_z - \tfrac{i}{2}(1-\ol{\zeta}^2) \pd_{\ol \zeta}\\
\end{aligned}
\ee
In particular, we see that they act by a diagonal action on $\PP^1_z \times \PP^1_\zeta$, with the coordinates $x^4$ and $r$ being invariant.

It is useful to work with projective coordinates on these two copies of $\PP^1$, making the above action linear.
On $\PP^1_z$ we have projective coordinates $\lambda_\alpha$, where $z = \lambda_2/\lambda_1$, and on $\PP^1_\zeta$ we introduce $\nu_\alpha$, where $\zeta = \nu_2 / \nu_1$.
The above rotations act on $\lambda_\alpha$, $\nu_\alpha$ as spinors of $\text{Spin}(3) \simeq SU(2)$.
Rotation invariants can be constructed in the usual fashion by contracting these spinors with the Levi-Civita tensor, such as $\langle \lambda \nu \rangle = \epsilon^{\beta \alpha} \lambda_\alpha \nu_\beta$.
We also introduce the conjugate spinor $\hat{\nu}_\alpha = (\ol{\nu}_2, -\ol{\nu}_1)$ satisfying $\langle \nu \hat{\nu} \rangle = |\nu|^2$, another $SU(2)$ invariant.
For example, the sections $q_\pm$, $\wt{q}_\pm$ are realized as the restriction of the following to their respective domains of definition:
\be
	\mathbf{Q}_+ = -\sqrt{2r}\frac{\langle \lambda \hat{\nu}\rangle}{\langle \nu \hat{\nu}\rangle^{\scriptstyle{\frac{1}{2}}}} \qquad \mathbf{Q}_- = \sqrt{2r}\frac{\langle \lambda \nu\rangle}{\langle \nu \hat{\nu}\rangle^{\scriptstyle{\frac{1}{2}}}}
\ee
Namely, $q_\pm$ is obtained by restricting $\mathbf{Q}_\pm$ to the open set $\lambda_1 \neq 0$ and $\nu_1 \neq 0$ and rescaling $\lambda_\alpha \to \lambda_\alpha / \lambda_1$ and $\nu_\alpha \to \nu_\alpha/\nu_1$.

\subsection{Bootstrapping tree-level MHV amplitudes}
In this section we show that the tree-level MHV scattering amplitudes of \cite{ABMS2} can be realized from the celestial chiral algebra found in Section \ref{sec:dyonCCA}.
To do this, we use the perspective of \cite{CP22} that we can reformulate these amplitudes as correlation functions of the celestial chiral algebra in integrals of certain conformal blocks $\langle \Tr(B^2)(x)|$.
In the next subsection we will provide justification for the name of these blocks, \emph{cf.} \cite{BC}.

We now introduce the conformal block $\langle \Tr(B^2)(x)|$, depending on a point $x$ on spacetime, that we use to reproduce the above tree-level MHV amplitudes.
The key properties of correlation functions in this conformal block are:
\begin{enumerate}
	\item[1)] vanishing unless there are two exactly two $\wt{J}$ operators
	\item[2)] vanishing if they contain $J^{\text{min}}_a[r_1, r_2]$ or $\wt{J}^{\text{min}}_a[r_1, r_2]$ with $r_1 + r_2 > 0$
	\item[3)] equivariance with respect to spacetime symmetries preserving the worldline of the dyon, including translations along $\ell$, rotations around $\ell$, and overall scaling $x \to \rho x$
	\item[4)] regularity in the spacetime coordinate $x$
\end{enumerate}
The first two constraints are identical to those appearing in \cite{CP22} that characterize the conformal block corresponding to the operator $\Tr(B^2)$.
The main difference from the case studied in \cite{CP22} is a weakening of full $SO(4)$ equivariance%
\footnote{More precisely, the paper \cite{CP22} imposed $SO(4)$ \emph{invariance} as it only considered the conformal block at the $SO(4)$ fixed point $x = 0$.} %
of correlation functions to the $SO(3)$ subgroup preserving the dyon worldline.

We now show that there is a unique conformal block satisfying these properties.
The first two conditions are consistent because the chiral algebra has a (non-cohomological) grading where $\wt{J}$ is given $1$ and $J$ is given weight $0$ and the minimal generators with $r_1 + r_2 \geq 0$ form an ideal.

Now consider the 2-point function of $\wt{J}^{\text{min}}_{a_1}[0,0]$ and $\wt{J}^{\text{min}}_{a_2}[0,0]$.
Note that invariance under the torus generated by the zero mode of $J^{\text{min}}_0[0,0]$ ensures this 2-point function is only nonzero if $e_1 + e_2 = 0$, where $e_i = e_{a_i}$.
We don't have full $\fg$-invariance as the values of $m_i$ cannot be less than the minimal value $\max(0, -e_i)$ -- we only retain the subalgebra of $\fg$ spanned by the generators $T_a$ with non-positive $T_0$ weight $e_a \leq 0$ as these generators allow for $m_a = 0$.
These subalgebras always contains a Borel that can be used to constrain the form of the overall Lie algebraic dependence.%
\footnote{It is easy to show that a symmetric bilinear form on $\fsl_n$ that is invariant with respect to a Borel is necessarily a multiple of the Killing form.
It would be interesting to see if this extends to general simple Lie algebra $\fg$.} %
The 2-point function thus takes the form
\be
	\langle \Tr(B^2)(x)|\wt{J}_{a_1}^{\text{min}}[0,0](z_1) \wt{J}_{a_2}^{\text{min}}[0,0](z_2)\rangle = F^{e_1, e_2}(x; z_1, z_2) \Tr(T_{a_1} T_{a_2})\,.
\ee

Translation invariance along $\ell$ guarantees that this correlation function does not depend on $x^4$.
Rotation invariance around the worldline implies that $F$ can only depend on the invariants $r$, $\langle \lambda_1 \lambda_2\rangle$, $\langle \lambda_i \nu\rangle$, $\langle \lambda_i \hat{\nu}\rangle$, and $\langle \nu \hat{\nu}\rangle$ when expressed in homogeneous coordinates.
Moreover, the form of the coupling between $\CB_{\pm, a}$ and $\wt{J}_{\pm, a}[r_1, r_2]$ implies that $\wt{J}_{\pm, a}[r_1, r_2]$, and hence this correlation function, must transform as a section of the $e_i$th power of the twistor quadrille twisted by $\CO(2+2m_i)$ with respect to $\lambda_i$.
Finally, scaling invariance implies that this 2-point function must scale with weight $\rho^{m_1 + m_2}$ under $r \to \rho r$.
The general solution to these symmetry constraints takes the form
\be
	r^{m_1 + m_2} \bigg(\sqrt{r} \frac{\langle \lambda_1 \hat{\nu} \rangle}{\langle \nu \hat{\nu}\rangle^{\scriptstyle{\frac{1}{2}}}}\bigg)^{e_1} \bigg(\sqrt{r} \frac{\langle \lambda_2 \hat{\nu} \rangle}{\langle \nu \hat{\nu}\rangle^{\scriptstyle{\frac{1}{2}}}}\bigg)^{e_2} \langle \lambda_1 \lambda_2\rangle^a \bigg(\frac{\langle \lambda_1 \nu \rangle \langle \lambda_1 \hat{\nu} \rangle}{\langle \nu \hat{\nu}\rangle}\bigg)^{m_1 + 1 - \scriptstyle{\frac{a}{2}}} \bigg(\frac{\langle \lambda_2 \nu \rangle \langle \lambda_2 \hat{\nu} \rangle}{\langle \nu \hat{\nu}\rangle}\bigg)^{m_2 + 1 - \scriptstyle{\frac{a}{2}}}
\ee
The first factor accounts for scaling of $r$, the second two factors trivialize the quadrille bundles, and the form of the exponents in the last three terms ensure the trivialized section has homogeneity $2+2m_i$ in $\lambda_i$ required of a section of $\CO(2+2m_i)$.

The final constraint we require is that this 2-point function is regular in the spacetime point $x \in \R^4$ that labels the conformal block.
The main consequence of this regularity is that each factor of $\langle \lambda_i \nu\rangle \langle \nu \hat{\nu}\rangle^{-\scriptstyle{\frac{1}{2}}}$ or $\langle \lambda_i \hat{\nu}\rangle \langle \nu \hat{\nu}\rangle^{-\scriptstyle{\frac{1}{2}}}$ must be accompanied by a factor of $\sqrt{r}$.
Namely, the 2-point function must depend on $\nu$ and $\hat{\nu}$ through $\mathbf{Q}_\pm$ evaluated at each $\lambda_i$.
Additionally, the overall powers of $r$ and $\mathbf{Q}_{\pm, i}$ must be non-negative integers.
Together, these imply the parameter $a$ must satisfy $2 \leq a \leq 2 + 2 m_i$.
As we are considering minimal generators and their charges $e_i$ satisfy $e_1 + e_2 = 0$, it follows that either $m_1 = 0$ or $m_2 = 0$ and hence regularity uniquely fixes $a = 2$.
We see that these symmetry and regularity constraints force this 2-point function to be given (up to an overall numerical factor) in homogeneous coordinates by:
\be
	\langle \Tr(B^2)(x)|\wt{J}_{a_1}^{\text{min}}[0,0](\lambda_1) \wt{J}_{a_2}^{\text{min}}[0,0](\lambda_2)\rangle = \left( \prod_{i = 1}^2 \mathbf{Q}_{+, i}^{m_i + e_i} \mathbf{Q}_{-, i}^{m_i} \right) \langle \lambda_1 \lambda_2\rangle^2 \Tr(T_{a_1} T_{a_2}).
\ee

We now extend this $2$-point function to higher points.
Namely, we will show that the color-ordered $n$-point function takes the form
\be
\begin{aligned}
	& \langle \Tr(B^2)(x)|J^{\text{min}}_{a_1}[0,0](\lambda_1) \dots \wt{J}_{a_r}^{\text{min}}[0,0](\lambda_r) \dots  \wt{J}_{a_s}^{\text{min}}[0,0](\lambda_s) \dots J^{\text{min}}_{a_n}[0,0](\lambda_n)\rangle_{CO}\\
	& \qquad = \bigg(\prod_{i=1}^n \mathbf{Q}_{+,i}^{m_i + e_i}\mathbf{Q}_{-, i}^{m_i}\bigg) \frac{\langle \lambda_r \lambda_s\rangle^4}{\langle \lambda_1 \lambda_2 \rangle \dots \langle \lambda_n \lambda_1 \rangle} \Tr(T_{a_1} \dots T_{a_n})
\end{aligned}
\ee
by induction on $n$; the full $n$-point function is realized as a sum of permutations of this expression.
The case $n = 2$ is automatic, so now consider the $(n+1)$-point correlator.
This expression has poles at $z_{n+1} = z_1, \dots, z_n$ whose residues are known $n$-point functions.
The only poles that can contribute to the color-ordered correlator are those at $z_{n+1} = z_1$ and $z_{n+1} = z_n$; the residues at these poles are given by
\be
\begin{aligned}
	& \bigg(\prod_{i=1}^{n} \mathbf{Q}_{+,i}^{m_i + e_i} \mathbf{Q}_{-,i}^{m_i}\bigg) \frac{\langle \lambda_r \lambda_s\rangle^4}{\langle \lambda_1 \lambda_2 \rangle \dots \langle \lambda_n \lambda_1 \rangle} \Tr([T_{a_{n+1}},T_{a_1}] \dots T_{a_n})\\
	& \hspace{2cm} \times \mathbf{Q}_{+,1}^{m_{n+1} + e_{n+1}}\mathbf{Q}_{-,1}^{m_{n+1}}
\end{aligned}
\ee
and
\be
\begin{aligned}
	& \bigg(\prod_{i=1}^{n} \mathbf{Q}_{+,i}^{m_i + e_i} \mathbf{Q}_{-,i}^{m_i}\bigg) \frac{\langle \lambda_r \lambda_s\rangle^4}{\langle \lambda_1 \lambda_2 \rangle \dots \langle \lambda_n \lambda_1 \rangle} \Tr(T_{a_1} \dots [T_{a_{n+1}},T_{a_n}])\\
	& \hspace{2cm} \times \mathbf{Q}_{+,n}^{m_{n+1} + e_{n+1}}\mathbf{Q}_{-,n}^{m_{n+1}}
\end{aligned}
\ee
respectively.
The color-ordered correlation function thus takes the form
\be
	\bigg(\prod_{i=1}^{n} \mathbf{Q}_{+,i}^{m_i + e_i} \mathbf{Q}_{-,i}^{m_i}\bigg) \frac{\langle \lambda_r \lambda_s\rangle^4}{\langle \lambda_1 \lambda_2 \rangle \dots \langle \lambda_{n} \lambda_{n+1} \rangle \langle \lambda_{n+1} \lambda_1 \rangle} F(x, \lambda_i) \Tr(T_{a_1} \dots T_{a_n} T_{a_{n+1}})\,.
\ee
Spacetime symmetry constraints say that $F$ must be rotation equivariant, scale as $r^{m_{n+1} + \scriptstyle{\frac{1}{2}}e_{n+1}}$, homogeneous of degree $0$ in $\lambda_i$ for $i < n+1$, and it must transform as in the $-e_{n+1}$th power of the twistor quadrille twisted by $\CO(2m_i)$ with respect to $\lambda_{n+1}$.
Together with the above residues, this uniquely fixes
\be
	F(x,\lambda_i) = \mathbf{Q}_{+,n+1}^{m_{n+1} + e_{n+1}} \mathbf{Q}_{-,n+1}^{m_{n+1}}
\ee
as desired.

As the correlation functions involving $\wt{J}_{a}^{\text{min}}[r_1,r_2]$ or $\wt{J}_{a_2}^{\text{min}}[r_1, r_2]$ must vanish if $r_1 + r_2 > 0$, we find that the color-ordered $n$-point function of two operators dual to minimal quasi-momentum eigenstates takes nearly the same form as the above $n$-point function:
\be
\begin{aligned}
	& \langle \Tr(B^2)(x)|J^{\text{min}}_{a_1}[\tilde{\lambda}_1](\lambda_1) \dots \wt{J}_{a_r}^{\text{min}}[\tilde{\lambda}_r](\lambda_r) \dots  \wt{J}_{a_s}^{\text{min}}[\tilde{\lambda}_s](\lambda_s) \dots J^{\text{min}}_{a_n}[\tilde{\lambda}_n](\lambda_n)\rangle_{CO}\\
	& \qquad = \bigg(\prod_{i=1}^n \mathbf{Q}_{+,i}^{m_i + e_i}\mathbf{Q}_{-, i}^{m_i} e^{i k_i \cdot x}\bigg) \frac{\langle \lambda_r \lambda_s\rangle^4}{\langle \lambda_1 \lambda_2 \rangle \dots \langle \lambda_n \lambda_1 \rangle} \Tr(T_{a_1} \dots T_{a_n})
\end{aligned}
\ee
where $(k_i)_{\alpha \dot \alpha} = -i \lambda_{i,\alpha} \tilde{\lambda}_{i, \dot \alpha}$ and $m_i = \max(0, -e_{a_i})$.
The additional plane-wave factors come from the definition of these operators in terms of the conformally soft modes.
We can differentiate these with (color-ordered) correlation functions as in Sections \ref{sec:quasimomentumstates} to get the (color-ordered) correlators of non-minimal generators with $m_i \geq \max(0, -e_{a_i})$ as well, leading to the following:
\be
\label{eq:dyonMHV}
\begin{aligned}
	& \langle \Tr(B^2)(x)|J^{(m_1)}_{a_1}[\tilde{\lambda}_1](\lambda_1) \dots \wt{J}_{a_r}^{(m_r)}[\tilde{\lambda}_r](\lambda_r) \dots  \wt{J}_{a_s}^{(m_s)}[\tilde{\lambda}_s](\lambda_s) \dots J^{(m_n)}_{a_n}[\tilde{\lambda}_n](\lambda_n)\rangle_{CO}\\
	& \qquad = \bigg(\prod_{i=1}^n \mathbf{Q}_{+,i}^{m_i + e_i}\mathbf{Q}_{-, i}^{m_i} e^{i k_i \cdot x}\bigg) \frac{\langle \lambda_r \lambda_s\rangle^4}{\langle \lambda_1 \lambda_2 \rangle \dots \langle \lambda_n \lambda_1 \rangle} \Tr(T_{a_1} \dots T_{a_n})
\end{aligned}
\ee

It is useful to express these correlation functions in terms of the affine coordinates, \emph{e.g.} on the patch with $\lambda_{i,1} \neq 0$ and $\nu_1 \neq 0$:
\be
\begin{aligned}
	\langle \Tr(B^2)(x)|J^{(m_1)}_{a_1}[\tilde{\lambda}_1](z_1) \dots \wt{J}_{a_r}^{(m_r)}[\tilde{\lambda}_r](z_r) \dots  \wt{J}_{a_s}^{(m_s)}[\tilde{\lambda}_s](z_s) \dots J^{(m_n)}_{a_n}[\tilde{\lambda}_n](z_n)\rangle_{CO}\\
	= \bigg(\prod_{i=1}^n q_+(x,z_i)^{m_i + e_i}q_-(x,z_i)^{m_i} e^{i k_i \cdot x}\bigg) \frac{(z_r - z_s)^4}{(z_1 - z_2) \dots (z_n - z_1)} \Tr(T_{a_1} \dots T_{a_n}) 
\end{aligned}
\ee
We interpret this $n$-point correlation function as a scattering amplitude in self-dual gauge theory in the presence of a self-dual dyon and together with an insertion of the operator $\Tr(B^2)$ at the spacetime point $x$.
If we integrate this expression over $x$, this should reproduce the tree-level $n$-point MHV scattering gluons in Yang-Mills theory.
Indeed, performing this integral precisely reproduces the recent results of \cite{ABMS1, ABMS2}.

\subsection{Tree-level MHV scattering from lifting \texorpdfstring{$\Tr(B^2)$}{TrB2} to \texorpdfstring{$\PT_Q$}{PTQ}}
Having shown that the scattering amplitudes of \cite{ABMS1, ABMS2} can be obtained from correlation functions of the celestial chiral algebra introduced in Section \ref{sec:dyonCCA} in a conformal block denoted $\langle \Tr(B^2)(x)|$, we now show why this is a suitable name for this block.
Namely, following \cite{BC}, we extract these correlators from lifting the 4d operator $\Tr(B^2)$ to $\PT_Q$ with a background Coulomb bundle.
Without the dyon, the lift of this composite operator, placed at a point $x \in \R^4$, to twistor space involves placing two insertions of $\CB_a$ at points $z_1, z_2$ on the twistor sphere over $x$, connecting them by a holomorphic Wilson line, and then integrating over these insertion points \cite{Mason:2005zm}.
The same is still true in the presence of the dyon, but we must take into account that the fields take values in (powers of) the Coulomb bundle by trivializing the Coulomb bundle on each twistor sphere (away from the worldline $\ell$) as above.

\subsubsection{Lifting \texorpdfstring{$\Tr(B^2)$}{TrB2} with the Coulomb bundle}
We first recall aspects of the lift of $\Tr(B^2)$ in the absence of the dyon and then describe how to account for the Coulomb bundle.
Let $\CA$ be a gauge field on $\PP^1$ (e.g. the restriction of the 6d gauge field to a twistor sphere) and let $z_1, z_2$ be points on that $\PP^1$.
The holomorphic Wilson line $W[z_1, z_2]$ from $z_1$ to $z_2$ is defined by the holomorphic parallel transport operator
\be
	W[z_1, z_2] = \text{Pexp}\bigg(\int_{\PP^1} \omega_{z_1, z_2} \CA\bigg) \qquad \omega_{z_1, z_2} = \frac{1}{\pi} \frac{(z_1 - z_2) \text{d}\sigma}{(z_1 - \sigma)(\sigma-z_2)}
\ee
where the path-ordered exponential is defined as 
\be
	\text{Pexp}\bigg(\int_{\PP^1} \omega_{z_1, z_2} \CA\bigg) = 1 + \sum_{n > 0} \int_{(\PP^1)^{n}} \omega^{(n)}_{z_1, z_2} \CA_1 \dots \CA_n\,,
\ee
where $\CA_i = \CA(\sigma_i)$ and
\be
	\omega^{(n)}_{z_1, z_2} = \frac{1}{\pi^n}\frac{(z_1 - z_2) \text{d} \sigma_1 \dots \text{d}\sigma_n}{(z_1 - \sigma_1)(\sigma_1 - \sigma_2) \dots (\sigma_n - z_2)}\,.
\ee
That this holomorphic Wilson line transforms as an ordinary Wilson line is due to the relation
\be
	\ol{\pd} \omega_{z_1, z_2} = \delta^2(\sigma - z_1) - \delta^2(\sigma - z_2)
\ee
and its higher-order analogues, which lead to the identities
\be
	\CQ W[z_1, z_2] = c(z_1) W[z_1, z_2] - W[z_1, z_2] c(z_2)\,,
\ee
and
\be
	\ol{\pd} W[z_1, z_2] + A(z_1) W[z_1, z_2] - W[z_1, z_2] A(z_2) = 0\,,
\ee
where $c = \CA^{(0)}$ and $A = \CA^{(1)}$ as above.
We can now describe the lift of the operator $\Tr(B^2)$ to twistor space: we place two insertions of $\CB$ on the twistor sphere $\PP^1$ over a point $x$, connect them by holomorphic Wilson lines, and then integrate over the twistor sphere with a suitable prefactor.
Explicitly, this lift takes the following form \cite{Mason:2005zm}:
\be
	\Tr(B^2) \text{ at } x \longleftrightarrow \int_{\PP^1 \times \PP^1} \text{d} z_1 \text{d} z_2 (z_1 - z_2)^2 \Tr(\CB(x, z_1) W[z_1, z_2] \CB(x, z_2) W[z_2, z_1])
\ee

There are two changes needed to lift this construction to $\PT_Q$.
We first describe an abelian Wilson line coupled to a background $\CG_0 = g_0$ that factorizes on each twistor sphere and we denote corresponding the trivialized gauge fields $\widehat{\CA}^H = (\CA_\pm^H, \CA_0^H)$.
The natural abelian expression takes the form
\be
\begin{aligned}
	W_\gamma[z_1, z_2] & = \exp\left(\int_{\PP^1, \gamma} \omega_{z_1, z_2} \widehat{\CA}^H\right)\\
	& := \exp\left(\int_{D_+} \omega_{z_1, z_2} \CA_+^H + \int_{D_-} \omega_{z_1, z_2} \CA_-^H + \int_{\gamma} \omega_{z_1, z_2} \CA_0^H\right)
\end{aligned}
\ee
We must be careful about how to understand the integral over $\gamma$ when either of $z_1$ or $z_2$ is on $\gamma$ due to the poles in $\omega_{z_1, z_2}$; in these cases, we define the integral by infinitesimally deforming $\gamma$ slightly so that the singularity lies in $D_+$.
This operator transforms covariantly with respect to gauge transformations, but we must keep track of where each of the endpoints lies:
\be
	\CQ W_\gamma[z_1, z_2] = \begin{cases}
		\big(c_+(z_1) - c_+(z_2)\big) W_\gamma[z_1, z_2] & z_1 \in D_+\,, ~ z_2 \in D_+\\
		\big(c_+(z_1) - c_-(z_2)\big) W_\gamma[z_1, z_2] & z_1 \in D_+\,, ~ z_2 \in \overset{\circ}{D_-}\\
		\big(c_-(z_1) - c_+(z_2)\big) W_\gamma[z_1, z_2] & z_1 \in \overset{\circ}{D_-}\,, ~ z_2 \in D_+\\
		\big(c_-(z_1) - c_-(z_2)\big) W_\gamma[z_1, z_2] & z_1 \in \overset{\circ}{D_-}\,, ~ z_2 \in \overset{\circ}{D_-}\\
	\end{cases}
\ee
where $c_\pm = \CA^{(0)}_\pm$ and $\overset{\circ}{D_-} = D_- \backslash \pd D_-$ is the interior of the $D_-$.
The dependence of this operator on the curve $\gamma$ is also very mild:
\be
\begin{aligned}
	W_{\gamma}[z_1, z_2] & = \exp\left(\int_{D_+ - D_+'} \omega_{z_1, z_2}(\CA^H_+ - \CA^H_-) + \int_{\gamma - \gamma'} \omega_{z_1, z_2} \CA^H_0\right) W_{\gamma'}[z_1, z_2]\\
	& = \exp\left(\int_{D_+ - D_+'} \ol{\pd}\omega_{z_1, z_2} \CA^H_0 - \CQ \left(\int_{D_+ - D_+'} \omega_{z_1, z_2} \CA^H_0\right) \right) W_\gamma[z_1, z_2]
\end{aligned}
\ee
This prefactor is $\CQ$-exact unless $z_1$ or $z_2$ belongs to $D_+ - D_+'$, \emph{i.e.} if $\gamma$ must cross $z_1$ or $z_2$ to get to $\gamma'$.
In particular, the different choices for dealing with cases where $z_1$ or $z_2$ lies on $\gamma$, such as deform $\gamma$ towards $D_+$ rather than $D_-$, can be easily related to each other; the factors of $\exp(\pm\CA_0^H(z_i)) = 1 \pm \CA_0^H(z_i) + \dots$ that appear when crossing $z_1$, $z_2$ ensure the consistency of gauge transformations for the different choices.
The four Wilson lines of interest will be denoted $W^{\varepsilon_1 \varepsilon_2}_\gamma[z_1, z_2]$, $\varepsilon_i = \pm$ denotes that we deform $\gamma$ away from $D_{\varepsilon_i}$ at $z_i$.
When $z_1$ is not equal to $\zeta$ or $-\ol{\zeta}^{-1}$, these Wilson lines are related as
\be
\label{eq:WLsign1}
	W^{+ \varepsilon_2}_\gamma[z_1, z_2] = (1 + \CA_0^H(z_1) + \dots) W^{- \varepsilon_2}_\gamma[z_1, z_2]
\ee
and when $z_2$ is not equal to $\zeta$ or $-\ol{\zeta}^{-1}$, these Wilson lines are related as
\be
\label{eq:WLsign2}
	W^{\varepsilon_1 +}_\gamma[z_1, z_2] = W^{\varepsilon_1 -}_\gamma[z_1, z_2](1 - \CA_0^H(z_2) + \dots)
\ee

The non-abelian holomorphic Wilson is realized by a path-ordered version of this construction, schematically of the following form:
\be
\begin{aligned}
	W_\gamma[z_1, z_2] & = \text{Pexp}\left(\int_{\PP^1, \gamma} \omega_{z_1, z_2} \widehat{\CA}\right) := 1 + \sum_{n > 0} \int_{(\PP^1, \gamma)^n} \omega^{(n)}_{z_1, z_2} \widehat{\CA}^H_1 \dots \widehat{\CA}^H_n\,.
\end{aligned}
\ee
The higher-order integrals should be interpreted as a sum of integrals over products of the chains $D_\pm$, $\gamma$ of the corresponding component fields $\CA_\pm$, $\CA_0$, which we do not write to avoid undue clutter.
As in the abelian setting, we must be careful about how to define the integral when the poles of $\omega^{(n)}_{z_1, z_2}$ are on $\gamma$.
The $(n,0)$-form $\omega^{(n)}_{z_1, z_2}$ has simple poles along the adjacent diagonals $\sigma_i = \sigma_{i+1}$ ($i = 1, \dots, n-1$) and when $\sigma_1 = z_1$ or $\sigma_n = z_2$.
There is a bit of flexibility for the poles on the adjacent diagonals and we define the integrals by taking the principal value, \emph{i.e.} we take the average of both options.
We deal with the latter poles as in the abelian setting by introducing the four variants $W^{\varepsilon_1 \varepsilon_2}_\gamma[z_1, z_2]$ where we define the integral by slightly deforming the first (resp. $n$th) copy of $\gamma$ so that this pole always lies in $D_{\varepsilon_1}$ (resp. $D_{\varepsilon_2}$).

When $z_1$ (resp. $z_2$) is neither $\zeta$ nor $-\ol{\zeta}^{-1}$, the Wilson lines $W^{+ \varepsilon_2}_\gamma[z_1, z_2]$ and $W^{- \varepsilon_2}_\gamma[z_1, z_2]$ (resp. $W^{\varepsilon_1 +}_\gamma[z_1, z_2]$ and $W^{\varepsilon_2 -}_\gamma[z_1, z_2]$) are related to one another as in Eq. \eqref{eq:WLsign1} (resp. Eq. \eqref{eq:WLsign2}). 
These non-abelian Wilson lines transform under gauge transform by conjugations depending on $\varepsilon_i$ and $z_i$:
\be
\label{eq:WLgaugetransform}
	\CQ W^{\varepsilon_1 \varepsilon_2}_\gamma[z_1, z_2] = \begin{cases}
		c_{\varepsilon_1}(z_1)^H W^{\varepsilon_1 \varepsilon_2}_\gamma[z_1, z_2] - W^{\varepsilon_1 \varepsilon_2}_\gamma[z_1, z_2] c_{\varepsilon_2}(z_2)^H & z_1 \in D_{\varepsilon_1}\,, ~ z_2 \in D_{\varepsilon_2}\\
		c_{\varepsilon_1}(z_1)^H W^{\varepsilon_1 \varepsilon_2}_\gamma[z_1, z_2] - W^{\varepsilon_1 \varepsilon_2}_\gamma[z_1, z_2] c_{-\varepsilon_2}(z_2)^H & z_1 \in D_{\varepsilon_1}\,, ~ z_2 \in \overset{\circ}{D_{-\varepsilon_2}}\\
		c_{-\varepsilon_1}(z_1)^H W^{\varepsilon_1 \varepsilon_2}_\gamma[z_1, z_2] - W^{\varepsilon_1 \varepsilon_2}_\gamma[z_1, z_2] c_{\varepsilon_2}(z_2)^H & z_1 \in \overset{\circ}{D_{-\varepsilon_1}}\,, ~ z_2 \in D_{\varepsilon_2}\\
		c_{-\varepsilon_1}(z_1)^H W^{\varepsilon_1 \varepsilon_2}_\gamma[z_1, z_2] - W^{\varepsilon_1 \varepsilon_2}_\gamma[z_1, z_2] c_{-\varepsilon_2}(z_2)^H & z_1 \in \overset{\circ}{D_{-\varepsilon_1}}\,, ~ z_2 \in \overset{\circ}{D_{-\varepsilon_2}}\\
	\end{cases}
\ee
That these tranformation properties are insensitive to the details of how we deal with the adjacent diagonals is ultimately due to the constraint in Eq. \eqref{eq:A0constraint}.

We are thus led to the following schematic expression for the lift of $\Tr(B^2)$ to $\PT_Q$ in the presence of the Coulomb bundle: 
\be
	\int_{(\PP^1_{Q,x}, \gamma) \times (\PP^1_{Q,x}, \gamma)} \text{d} z_1 \text{d} z_2 (z_1 - z_2)^2 \Tr\left(\widehat{\CB}^H(x, z_1) W_\gamma[z_1, z_2] \widehat{\CB}^H(x, z_2) W_{\gamma}[z_2, z_1]\right)\\
\ee
This expression should be interpreted as a sum of integrals over products of the chains $D_\pm$, $\gamma$ of the corresponding component fields $\CB_\pm$, $\CB_0$ connected by the Wilson lines $W^{\varepsilon_1 \varepsilon_2}_\gamma[z_1, z_2]$ depending on the chain, where $\varepsilon_i = +$ if $z_i$ is integrated over $D_+$ or $\gamma$ and $\varepsilon_i = -$ if $z_i$ is integrated over $D_-$.

\subsubsection{Amplitudes from the lift of \texorpdfstring{$\Tr(B^2)$}{TrB2}}

We can extract tree-level MHV scattering amplitudes from the above expressions for the lift of $\Tr(B^2)$.
To get the $(n+2)$-point MHV amplitude, we consider the terms in the lift of $\Tr(B^2)$ proportional to $n$ factors of $\CA$.
In the absence of the dyon, the terms containing $n$ factors of $\CA$ and $2$ factors of $\CB$ can be realized by permutations of the color-ordered operator
\be
\begin{aligned}
	& \int_{(\PP^1_x)^{n+2}} \text{d}\sigma_1 \dots \text{d} \sigma_{n+2} \frac{(\sigma_i - \sigma_j)^4}{(\sigma_1 - \sigma_2) \dots (\sigma_{n+2} - \sigma_1)}\\
	& \qquad \times \Tr(\dots \CA(x, \sigma_{i-1}) \CB(x, \sigma_{i}) \CA(x, \sigma_{i+1}) \dots \CA(x, \sigma_{j-1}) \CB(x, \sigma_{j}) \CA(x, \sigma_{j+1}) \dots)
\end{aligned}\,.
\ee
The color-ordered amplitude can be extracted by inserting plane-wave states with each copy of $\CA$ or $\CB$ of the form $\delta(z - z_i) \exp(v \cdot \tilde{\lambda}_i) T_{a_i}$ into this expression and then integrating over the insertion point $x$:
\be
	\CA^{CO}_{n+2} = \int \text{d}^4 x \frac{(z_i - z_j)^4}{(z_1 - z_2) \dots (z_{n+2} - z_{1})} \Tr(T_{a_1} \dots T_{a_{n+2}}) e^{i k \cdot x}
\ee
where $k = \sum_i k_i$.

We can similarly extract the form factors associated to scattering quasi-momentum eigenstates off of the operator $\Tr(B^2)$ in the presence of a background self-dual dyon.
Indeed, when restricted to states with $\CA_0 = 0$ and $\CB_0 = 0$, the lift of $\Tr(B^2)$ to $\PT_Q$ is precisely the integrand of the MHV generating functional of \cite{ABMS2}.
The 2-point form factor is deduced from the form of the $\widehat{\CA}$-independent term in the lift of $\Tr(B^2)$:
\be
	\int_{(\PP^1_{Q,x}, \gamma) \times (\PP^1_{Q,x}, \gamma)} \text{d} \sigma_1 \text{d} \sigma_2 (\sigma_1 - \sigma_2)^2 \Tr(\widehat{\CB}^H(x, \sigma_1) \widehat{\CB}^H(x, \sigma_2))
\ee
Inserting the quasi-momentum eigenstates and integrating over $x$ leads the 2-point amplitude (up to overall factors)
\be
\label{eq:dyon2ptfcns}
	\CA_2 = \int_x \text{d}^4x ~ \bigg(\prod_{i=1,2} q_+(x,z_i)^{m_i+e_{a_i}} q_-(x,z_i)^{m_i} e^{i k_i \cdot x} \bigg) (z_1-z_2)^2 \Tr(T_{a_1} T_{a_2})
\ee
in agreement with \cite{ABMS1}.
We can similarly extract the $n$-point form factor of 2 negative helicity and $n-2$ positive helicity quasi-momentum eigenstates.
We find the color-ordered amplitude is given by (up to numerical factors)
\be
\label{eq:dyoncorrfcns}
	\CA^\text{CO}_{n} = \int_x \text{d}^4 x ~ \bigg(\prod_{i=1}^n q_+(x,z_i)^{m_i+e_{a_i}} q_-(x,z_i)^{m_i} e^{i k_i \cdot x} \bigg) \frac{(z_r-z_s)^4}{(z_1 - z_2) \dots (z_{n} - z_1)} \Tr(T_{a_1} \dots T_{a_n})
\ee
in agreement with \cite{ABMS2}.

\section{Acknowledgments}
We thank A. Sharma and L. Mason for very useful explanations of the twistor quadrille and K. Costello and T. Adamo for comments on a draft. 
NG and NP are supported by funds from the Department of Physics and the College of Arts \& Sciences at the University of Washington. NP also acknowledges support from the DOE Early Career Research Program under award DE-SC0022924, and the Simons Foundation as part of the Simons Collaboration on Celestial Holography.

\begin{appendix}

\section{The twistor quadrille}
\label{sec:quadrille}
The classic approach to the twistorial realization of the self-dual dyon is in terms of a line bundle, called the twistor quadrille, on the space of twistor spheres away from the worldline $\ell$ \cite{PS79}; see also \cite{ABMS2} for a recent treatment that we follow.

The construction of the twistor quadrille starts by using the identification $\PT \simeq \PP^1 \times \R^4$ in Euclidean signature, which is realized by inverting the incidence relation:
\be
	x^{\alpha \dot \alpha} = \frac{1}{1+|z|^2}\begin{pmatrix}
		\ol{v}^{\dot 2} + \ol{z} v^{\dot 1} & -\ol{v}^{\dot 1} + \ol{z} v^{\dot 2}\\
		v^{\dot 1} - z \ol{v}^{\dot 2} & v^{\dot 2} + z \ol{v}^{\dot 1}\\
	\end{pmatrix}
\ee
We can cover the complement of the wordline $\R^4 \backslash \ell \simeq (\R^3 \backslash \{0\}) \times \R$ with two open patches using stereographic coordinates on the spheres linking the dyon (the spheres around $0$ in $\R^3$).
Away from the positive $x^3$ axis, we choose coordinates on $\R^3 \backslash \{0\}$ given by
\be
	r = \sqrt{(x^1)^2 + (x^2)^2 + (x^3)^2} \qquad \zeta = \frac{x^1 + i x^2}{r - x^3} \qquad \ol{\zeta} = \frac{x^1 - i x^2}{r - x^3}\,,
\ee
and away from the negative $x^3$ axis, we choose coordinates
\be
	r = \sqrt{(x^1)^2 + (x^2)^2 + (x^3)^2} \qquad \omega = \frac{x^1 - i x^2}{r + x^3} \qquad \ol{\omega} = \frac{x^1 + i x^1}{r + x^3}\,.
\ee
On the overlap of these patches, \emph{i.e.} away from the $x^3$ axis, the complex coordinates are related as $\zeta = \omega^{-1}$.
It is worth noting that $\zeta$ and $\omega$ are \emph{not} holomorphic with respect to the complex structure induced from $\PT$.

The key observation is that the holomorphic section of $\CO(2) \to \PT$ given by $Q = v^{\dot 1} + z v^{\dot 2}$ away from $w=0$ and $\wt{Q} = w u^{\dot 1} + u^{\dot 2}$ away from $z = 0$ can be factored upon restriction to the complement of the worldline $\ell$.
In particular, away from the negative $x^3$ axis on $\R^3\backslash\{0\}$ and away from $\lambda_1 = 0$ on $\PP^1$ we can write $Q = q_- q_+$ where $q_\pm$ are given by
\be
	q_-(z, \zeta) = \bigg(\frac{2r}{1+|\zeta|^2}\bigg)^{1/2}(z - \zeta) \qquad q_+(z, \zeta) = -\bigg(\frac{2r}{1+|\zeta|^2}\bigg)^{1/2}(1 + \ol{\zeta} z)\,,
\ee
Similarly, away from the positive $x^3$ axis we can write $Q = q'_+ q'_-$ for
\be
	q_-'(z, \omega) = \bigg(\frac{2r}{1+|\omega|^2}\bigg)^{1/2}(1 - \omega z) \qquad q_+'(z,\omega) = -\bigg(\frac{2r}{1+|\omega|^2}\bigg)^{1/2}(z + \ol{\omega})\,.
\ee
Away from the $x^3$ axis these are factors related by an overall phase
\be
	q_-(z, 1/\omega) = \frac{|\omega|}{\omega} q_-'(z, \omega) \qquad q_+(z, 1/\omega) = \frac{|\omega|}{\ol{\omega}} q_+'(z, \omega)
\ee
which indicates they are sections of the Wu-Yang monopole bundle on $S^2$ (with opposite charges).
A similar story goes through on the other patch of $\PP^1$. 
For example, away from the negative $x^3$ axis we can write $\wt{Q} = \wt{q}_- \wt{q}_+$ with 
\be
	\wt{q}_- = \bigg(\frac{2r}{1+|\zeta|^2}\bigg)^{1/2}(1 - \zeta w) \qquad \wt{q}_+ = \bigg(\frac{2r}{1+|\zeta|^2}\bigg)^{1/2}(w + \ol{\zeta})\,.
\ee

We now define a new cover of $\PP^1 \times (\R^4\backslash \ell)$ via these sections.
Focusing on points away from the negative $x^3$ axis for simplicity, we define the open sets $V_\pm$ by the condition that $\lambda_1 \neq 0$ and $q_\pm \neq 0$.
Similarly, we define the open sets $U_\pm$ by the condition that $\lambda_2 \neq 0$ and $\wt{q}_\pm \neq 0$.
(These 
With these open sets we can define transition functions for the twistor quadrille by positing that the $q_\pm^{\pm1}$ and $\wt{q}_\pm^{\pm1}$ serve to trivialize the bundle on $V_\pm$ and $U_\pm$, \emph{i.e.} if $s$ is a section of the quadrille and $s_{V_\pm}$, $s_{U_\pm}$ are its restrictions to these open sets then $q_\pm^{\pm1} s_{V_\pm}$ and  $\wt{q}_\pm^{\pm1} s_{U_\pm}$ define a section of the trivial bundle.
For example, on the overlap of $V_+$ and $V_-$ the transition function for the twistor quadrille is determined as
\be
	q^{-1}_- s_{V_-}\big|_{V_+ \cap V_-} = q_+ s_{V_+}\big|_{V_+ \cap V_-} \quad \rightsquigarrow \quad s_{V_-}\big|_{V_+ \cap V_-} = q_+ q_- s_{V_+}\big|_{V_+ \cap V_-} = Q s_{V_+}\big|_{V_+ \cap V_-}\,.
\ee
Similarly, on the overlap of $U_+$ and $V_+$ the transition function is
\be
	s_{U_+}\big|_{U_+ \cap V_+} = \frac{q_+}{\wt{q}_+} s_{V_+}\big|_{U_+ \cap V_+} = z s_{V_+}\big|_{U_+ \cap V_+}\,,
\ee
and on the overlap of $U_+$ and $V_-$ the transition function is
\be
	s_{U_+}\big|_{U_+ \cap V_-} = \frac{1}{q_-\wt{q}_+} s_{V_-}\big|_{U_+ \cap V_-} = \frac{z}{Q} s_{V_-}\big|_{U_+ \cap V_-}\,.
\ee
On the triple overlap of $U_+$, $V_+$, and $V_-$ we see that
\be
	s_{U_+}\big|_{U_+ \cap V_+ \cap V_-} = \frac{z}{Q} s_{V_-}\big|_{U_+ \cap V_+ \cap V_-} = \frac{z}{Q} \bigg(Q s_{V_+}\big|_{U_+ \cap V_+ \cap V_-}\bigg) =  z s_{V_+}\big|_{U_+ \cap V_+ \cap V_-}\,.
\ee
Importantly, although $\PP^1 \times (\R^4\backslash\ell)$ is not a complex manifold, all of the transition functions for the twistor quadrille are holomorphic when viewed as functions on the complex manifold $\PT$.
Indeed, as we describe in Section \ref{sec:coulomb}, this bundle over $\PP^1$ should rather be thought of as the restriction of the (genuinely holomorphic) Coulomb bundle $\CC \to \PT_Q$ to a twistor sphere.


\end{appendix}
\bibliographystyle{JHEP}
\bibliography{dyon}

\providecommand{\href}[2]{#2}\begingroup\raggedright\begin{thebibliography}{10}

\bibitem{rubakov1988monopole}
V.~Rubakov, \emph{Monopole catalysis of proton decay}, {\emph{Reports on
  Progress in Physics} {\bfseries 51} (1988) 189}.

\bibitem{callan1983monopole}
C.~G. Callan~Jr, \emph{Monopole catalysis of baryon decay}, {\emph{Nuclear
  Physics B} {\bfseries 212} (1983) 391}.

\bibitem{Rubakov:1982fp}
V.~A. Rubakov, \emph{{Adler-Bell-Jackiw Anomaly and Fermion Number Breaking in
  the Presence of a Magnetic Monopole}},
  \href{https://doi.org/10.1016/0550-3213(82)90034-7}{\emph{Nucl. Phys. B}
  {\bfseries 203} (1982) 311}.

\bibitem{Csaki:2021ozp}
C.~Cs\'aki, Y.~Shirman, O.~Telem and J.~Terning, \emph{{Pairwise Multiparticle
  States and the Monopole Unitarity Puzzle}},
  \href{https://doi.org/10.1103/PhysRevLett.129.181601}{\emph{Phys. Rev. Lett.}
  {\bfseries 129} (2022) 181601}
  [\href{https://arxiv.org/abs/2109.01145}{{\ttfamily 2109.01145}}].

\bibitem{Csaki:2022tvb}
C.~Cs\'aki, Z.-Y. Dong, O.~Telem, J.~Terning and S.~Yankielowicz,
  \emph{{Dressed vs. pairwise states, and the geometric phase of monopoles and
  charges}}, \href{https://doi.org/10.1007/JHEP02(2023)211}{\emph{JHEP}
  {\bfseries 02} (2023) 211}
  [\href{https://arxiv.org/abs/2209.03369}{{\ttfamily 2209.03369}}].

\bibitem{Hamada:2022eiv}
Y.~Hamada, T.~Kitahara and Y.~Sato, \emph{{Monopole-fermion scattering and
  varying Fock space}},
  \href{https://doi.org/10.1007/JHEP11(2022)116}{\emph{JHEP} {\bfseries 11}
  (2022) 116} [\href{https://arxiv.org/abs/2208.01052}{{\ttfamily
  2208.01052}}].

\bibitem{Brennan:2023tae}
T.~D. Brennan, \emph{{A new solution to the Callan Rubakov effect}},
  \href{https://doi.org/10.1007/JHEP11(2024)170}{\emph{JHEP} {\bfseries 11}
  (2024) 170} [\href{https://arxiv.org/abs/2309.00680}{{\ttfamily
  2309.00680}}].

\bibitem{brennan2023callan}
T.~D. Brennan, \emph{Callan-rubakov effect and higher charge monopoles},
  {\emph{Journal of High Energy Physics} {\bfseries 2023} (2023) 1}.

\bibitem{vanBeest:2023dbu}
M.~van Beest, P.~Boyle~Smith, D.~Delmastro, Z.~Komargodski and D.~Tong,
  \emph{{Monopoles, Scattering, and Generalized Symmetries}},
  \href{https://arxiv.org/abs/2306.07318}{{\ttfamily 2306.07318}}.

\bibitem{vanBeest:2023mbs}
M.~van Beest, P.~Boyle~Smith, D.~Delmastro, R.~Mouland and D.~Tong,
  \emph{{Fermion-monopole scattering in the Standard Model}},
  \href{https://doi.org/10.1007/JHEP08(2024)004}{\emph{JHEP} {\bfseries 08}
  (2024) 004} [\href{https://arxiv.org/abs/2312.17746}{{\ttfamily
  2312.17746}}].

\bibitem{Mouland:2024zgk}
R.~Mouland and D.~Tong, \emph{{On the Hilbert Space of Dyons}},
  \href{https://doi.org/10.1103/PhysRevD.110.085014}{\emph{Phys. Rev. D}
  {\bfseries 110} (2024) 085014}
  [\href{https://arxiv.org/abs/2401.01924}{{\ttfamily 2401.01924}}].

\bibitem{Costello:2021bah}
K.~J. Costello, \emph{{Quantizing local holomorphic field theories on twistor
  space}},  \href{https://arxiv.org/abs/2111.08879}{{\ttfamily 2111.08879}}.

\bibitem{CP22}
K.~Costello and N.~M. Paquette, \emph{{Celestial holography meets twisted
  holography: 4d amplitudes from chiral correlators}},
  \href{https://doi.org/10.1007/JHEP10(2022)193}{\emph{JHEP} {\bfseries 10}
  (2022) 193} [\href{https://arxiv.org/abs/2201.02595}{{\ttfamily
  2201.02595}}].

\bibitem{bailey1985twistors}
T.~Bailey, \emph{Twistors and fields with sources on worldlines},
  {\emph{Proceedings of the Royal Society of London. A. Mathematical and
  Physical Sciences} {\bfseries 397} (1985) 143}.

\bibitem{Bailey:1989if}
T.~N. Bailey and M.~A. Singer, \emph{{ON THE TWISTOR DESCRIPTION OF SOURCED
  FIELDS}}, \href{https://doi.org/10.1098/rspa.1989.0035}{\emph{Proc. Roy. Soc.
  Lond. A} {\bfseries 422} (1989) 367}.

\bibitem{MH1990}
L.~Mason and L.~Hughston, \emph{Further Advances in Twistor Theory}, no.~v. 1
  in Chapman \& Hall/CRC research notes in mathematics series. Longman
  Scientific \& Technical, 1990.

\bibitem{Costello:2019jsy}
K.~Costello and S.~Li, \emph{{Anomaly cancellation in the topological string}},
  \href{https://doi.org/10.4310/ATMP.2020.v24.n7.a2}{\emph{Adv. Theor. Math.
  Phys.} {\bfseries 24} (2020) 1723}
  [\href{https://arxiv.org/abs/1905.09269}{{\ttfamily 1905.09269}}].

\bibitem{PS79}
R.~Penrose and G.~Sparling, \emph{{The Twistor Quadrille: A Line Bundle Based
  on the Coulomb Field}},  in \emph{Advances in Twistor Theory}, L.~J. Mason,
  L.~P. Hughston, P.~Z. Kobak and K.~Pulverer, eds., CRC Press, (1979).

\bibitem{PS90}
R.~Penrose and G.~Sparling, \emph{{The anti-self-dual Coulomb field's
  non-Hausdorff twistor space}},  in \emph{Further Advances in Twistor Theory},
  L.~J. Mason and L.~P. Hughston, eds., vol.~1, Longman Scientific \&
  Technical, (1990).

\bibitem{ABMS1}
T.~Adamo, G.~Bogna, L.~Mason and A.~Sharma, \emph{{Scattering on self-dual
  Taub-NUT}}, \href{https://doi.org/10.1088/1361-6382/ad12ee}{\emph{Class.
  Quant. Grav.} {\bfseries 41} (2024) 015030}
  [\href{https://arxiv.org/abs/2309.03834}{{\ttfamily 2309.03834}}].

\bibitem{ABMS2}
T.~Adamo, G.~Bogna, L.~Mason and A.~Sharma, \emph{{Gluon scattering on the
  self-dual dyon}},
  \href{https://doi.org/10.1007/s11005-025-01907-2}{\emph{Lett. Math. Phys.}
  {\bfseries 115} (2025) 18}
  [\href{https://arxiv.org/abs/2406.09165}{{\ttfamily 2406.09165}}].

\bibitem{Adamo:2020yzi}
T.~Adamo, L.~Mason and A.~Sharma, \emph{{Gluon Scattering on Self-Dual
  Radiative Gauge Fields}},
  \href{https://doi.org/10.1007/s00220-022-04582-9}{\emph{Commun. Math. Phys.}
  {\bfseries 399} (2023) 1731}
  [\href{https://arxiv.org/abs/2010.14996}{{\ttfamily 2010.14996}}].

\bibitem{Adamo:2022mev}
T.~Adamo, L.~Mason and A.~Sharma, \emph{{Graviton scattering in self-dual
  radiative space-times}},
  \href{https://doi.org/10.1088/1361-6382/acc233}{\emph{Class. Quant. Grav.}
  {\bfseries 40} (2023) 095002}
  [\href{https://arxiv.org/abs/2203.02238}{{\ttfamily 2203.02238}}].

\bibitem{ABZ23}
T.~Adamo, W.~Bu and B.~Zhu, \emph{{Infrared structures of scattering on
  self-dual radiative backgrounds}},
  \href{https://doi.org/10.1007/JHEP06(2024)076}{\emph{JHEP} {\bfseries 06}
  (2024) 076} [\href{https://arxiv.org/abs/2309.01810}{{\ttfamily
  2309.01810}}].

\bibitem{Guevara:2021abz}
A.~Guevara, E.~Himwich, M.~Pate and A.~Strominger, \emph{{Holographic symmetry
  algebras for gauge theory and gravity}},
  \href{https://doi.org/10.1007/JHEP11(2021)152}{\emph{JHEP} {\bfseries 11}
  (2021) 152} [\href{https://arxiv.org/abs/2103.03961}{{\ttfamily
  2103.03961}}].

\bibitem{Strominger:2021mtt}
A.~Strominger, \emph{{$w_{1+\infty}$ Algebra and the Celestial Sphere: Infinite
  Towers of Soft Graviton, Photon, and Gluon Symmetries}},
  \href{https://doi.org/10.1103/PhysRevLett.127.221601}{\emph{Phys. Rev. Lett.}
  {\bfseries 127} (2021) 221601}
  [\href{https://arxiv.org/abs/2105.14346}{{\ttfamily 2105.14346}}].

\bibitem{Costello:2022upu}
K.~Costello and N.~M. Paquette, \emph{{Associativity of One-Loop Corrections to
  the Celestial Operator Product Expansion}},
  \href{https://doi.org/10.1103/PhysRevLett.129.231604}{\emph{Phys. Rev. Lett.}
  {\bfseries 129} (2022) 231604}
  [\href{https://arxiv.org/abs/2204.05301}{{\ttfamily 2204.05301}}].

\bibitem{Costello:2023vyy}
K.~J. Costello, \emph{{Bootstrapping two-loop QCD amplitudes}},
  \href{https://arxiv.org/abs/2302.00770}{{\ttfamily 2302.00770}}.

\bibitem{Dixon:2024mzh}
L.~J. Dixon and A.~Morales, \emph{{On gauge amplitudes first appearing at two
  loops}}, \href{https://doi.org/10.1007/JHEP08(2024)129}{\emph{JHEP}
  {\bfseries 08} (2024) 129}
  [\href{https://arxiv.org/abs/2407.13967}{{\ttfamily 2407.13967}}].

\bibitem{CP20}
K.~Costello and N.~M. Paquette, \emph{{Twisted Supergravity and Koszul Duality:
  A case study in AdS$_3$}},
  \href{https://doi.org/10.1007/s00220-021-04065-3}{\emph{Commun. Math. Phys.}
  {\bfseries 384} (2021) 279}
  [\href{https://arxiv.org/abs/2001.02177}{{\ttfamily 2001.02177}}].

\bibitem{PW21}
N.~M. Paquette and B.~R. Williams, \emph{{Koszul duality in quantum field
  theory.}}, \href{https://doi.org/10.5802/cml.88}{\emph{Confluentes Math.}
  {\bfseries 14} (2023) 87} [\href{https://arxiv.org/abs/2110.10257}{{\ttfamily
  2110.10257}}].

\bibitem{Bailey90}
T.~N. Bailey, \emph{{Why relative cohomology describes sources}},  in
  \emph{Further Advances in Twistor Theory}, L.~J. Mason and L.~P. Hughston,
  eds., vol.~1, Longman Scientific \& Technical, (1990).

\bibitem{Bittleston:2022jeq}
R.~Bittleston, \emph{{On the associativity of 1-loop corrections to the
  celestial operator product in gravity}},
  \href{https://doi.org/10.1007/JHEP01(2023)018}{\emph{JHEP} {\bfseries 01}
  (2023) 018} [\href{https://arxiv.org/abs/2211.06417}{{\ttfamily
  2211.06417}}].

\bibitem{Garner:2023izn}
N.~Garner and N.~M. Paquette, \emph{{Twistorial monopoles \& chiral algebras}},
  \href{https://doi.org/10.1007/JHEP08(2023)088}{\emph{JHEP} {\bfseries 08}
  (2023) 088} [\href{https://arxiv.org/abs/2305.00049}{{\ttfamily
  2305.00049}}].

\bibitem{CPS1}
K.~Costello, N.~M. Paquette and A.~Sharma, \emph{{Top-Down Holography in an
  Asymptotically Flat Spacetime}},
  \href{https://doi.org/10.1103/PhysRevLett.130.061602}{\emph{Phys. Rev. Lett.}
  {\bfseries 130} (2023) 061602}
  [\href{https://arxiv.org/abs/2208.14233}{{\ttfamily 2208.14233}}].

\bibitem{CPS2}
K.~Costello, N.~M. Paquette and A.~Sharma, \emph{{Burns space and holography}},
  \href{https://doi.org/10.1007/JHEP10(2023)174}{\emph{JHEP} {\bfseries 10}
  (2023) 174} [\href{https://arxiv.org/abs/2306.00940}{{\ttfamily
  2306.00940}}].

\bibitem{Bittleston:2023bzp}
R.~Bittleston, S.~Heuveline and D.~Skinner, \emph{{The celestial chiral algebra
  of self-dual gravity on Eguchi-Hanson space}},
  \href{https://doi.org/10.1007/JHEP09(2023)008}{\emph{JHEP} {\bfseries 09}
  (2023) 008} [\href{https://arxiv.org/abs/2305.09451}{{\ttfamily
  2305.09451}}].

\bibitem{Kapustin:2006pk}
A.~Kapustin and E.~Witten, \emph{{Electric-Magnetic Duality And The Geometric
  Langlands Program}},
  \href{https://doi.org/10.4310/CNTP.2007.v1.n1.a1}{\emph{Commun. Num. Theor.
  Phys.} {\bfseries 1} (2007) 1}
  [\href{https://arxiv.org/abs/hep-th/0604151}{{\ttfamily hep-th/0604151}}].

\bibitem{BottTu}
R.~Bott and W.~T. Loring, \emph{{Differential Forms in Algebraic Topology}},
  Graduate Texts in Mathematics. Springer New York, NY, 1982.

\bibitem{Adamo:2021zpw}
T.~Adamo, W.~Bu, E.~Casali and A.~Sharma, \emph{{Celestial operator products
  from the worldsheet}},
  \href{https://doi.org/10.1007/JHEP06(2022)052}{\emph{JHEP} {\bfseries 06}
  (2022) 052} [\href{https://arxiv.org/abs/2111.02279}{{\ttfamily
  2111.02279}}].

\bibitem{GO19}
D.~Gaiotto and J.~Oh, \emph{{Aspects of \ensuremath{\Omega}-deformed
  M-theory}}, \href{https://doi.org/10.1007/JHEP12(2024)184}{\emph{JHEP}
  {\bfseries 12} (2024) 184}
  [\href{https://arxiv.org/abs/1907.06495}{{\ttfamily 1907.06495}}].

\bibitem{BC}
W.~Bu and E.~Casali, \emph{{The 4d/2d correspondence in twistor space and
  holomorphic Wilson lines}},
  \href{https://doi.org/10.1007/JHEP11(2022)076}{\emph{JHEP} {\bfseries 11}
  (2022) 076} [\href{https://arxiv.org/abs/2208.06334}{{\ttfamily
  2208.06334}}].

\bibitem{Mason:2005zm}
L.~J. Mason, \emph{{Twistor actions for non-self-dual fields: A Derivation of
  twistor-string theory}},
  \href{https://doi.org/10.1088/1126-6708/2005/10/009}{\emph{JHEP} {\bfseries
  10} (2005) 009} [\href{https://arxiv.org/abs/hep-th/0507269}{{\ttfamily
  hep-th/0507269}}].

\end{thebibliography}\endgroup

\end{document}